\documentclass[12pt]{iopart}

\usepackage[english]{babel}
\usepackage{mathrsfs}
\expandafter\let\csname equation*\endcsname\relax
\expandafter\let\csname endequation*\endcsname\relax
\usepackage{amsmath}
\usepackage{amsthm}
\usepackage{amsfonts}
\usepackage{BOONDOX-cal}
\usepackage{graphicx}
\usepackage{braket}
\usepackage{subcaption}
\usepackage{multirow}
\usepackage[table]{xcolor}
\definecolor{nicepurple}{HTML}{97a0cf}
\definecolor{lightgreen}{HTML}{88BFB7} 
\definecolor{lightergreen}{HTML}{88BFB7} 
\definecolor{cgreen}{HTML}{88BFB7}
\definecolor{melon}{HTML}{FCBCB8}
\definecolor{puce}{HTML}{C08497}

\renewcommand{\newline}{\\\\\noindent}
\newcommand{\hilb}{\mathscr{H}}
\newcommand{\quotes}[1]{``#1"}

\newcommand{\extd}{\textrm{d}}
\newcommand{\Uone}{\textrm{U}(1)}
\newcommand{\Uqone}{\textrm{U}_q(1)}
\newcommand{\R}{\mathbb{R}}
\newcommand{\Z}{\mathbb{Z}}
\newcommand{\spacetime}{\mathcal{M}}
\newcommand{\jmax}{m_{max}}
\newcommand{\cnf}{\mathcal{T}}
\DeclareMathOperator{\sign}{sign}
\newcommand{\expect}[1]{{\langle #1\rangle}}

\usepackage{bbm}
\newcommand{\one}{\mathbbm{1}}

\usepackage{perpage}

\begin{document}

\title[]{Towards quantum gravity with neural networks: Solving quantum Hamilton constraints of 3d Euclidean gravity in the weak coupling limit}

\author{Hanno Sahlmann $^1$, Waleed Sherif $^2$}

\address{Institute for Quantum Gravity, Department of Physics, Friedrich-Alexander-Universit\"{a}t Erlangen-N\"{u}rnberg (FAU), Staudtstraße 7, 91058 Erlangen, Germany}
\ead{$^1$ hanno.sahlmann@fau.de, $^2$ waleed.sherif@gravity.fau.de}
\vspace{10pt}
\begin{indented}
\item[] \today
\end{indented}

\begin{abstract}
We consider 3-dimensional Euclidean gravity in the weak coupling limit of Smolin and show that it is BF-theory with $\Uone^3$ as a Lie group. The theory is quantised using loop quantum gravity methods. The kinematical degrees of freedom are truncated, on account of computational feasibility, by fixing a graph and deforming the algebra of the holonomies to impose a cutoff on the charge vectors. This leads to a quantum theory related to $\Uqone^3$ BF-theory. The effect of imposing the cutoff on the charges is examined. We also implement the quantum volume operator of 3d loop quantum gravity. Most importantly we compare two constraints for the quantum model obtained: a master constraint enforcing curvature and Gau{\ss} constraint, as well as a combination of a quantum Hamilton constraint constructed using Thiemann's strategy and the Gau{\ss} master constraint. The two constraints are solved using the neural network quantum state ansatz, demonstrating its ability to explore models which are out of reach for exact numerical methods. The solutions spaces are quantitatively compared and although the forms of the constraints are radically different, the solutions turn out to have a surprisingly large overlap. We also investigate the behavior of the quantum volume in solutions to the constraints. 
\end{abstract}

\section{Introduction}
The problem of finding a quantum theory of gravity is a long standing one with different attempts to resolving it \cite{Oriti:2009zz}. Einstein’s theory of general relativity (GR) describes an intricate interplay between the 4-dimensional spacetime and the matter within it \cite{Wald:1984rg}. The classical theory has proven to be remarkably successful, the direct observation of black holes \cite{EventHorizonTelescope:2019dse} and detection of gravitational waves \cite{LIGOScientific:2016aoc} serve as examples. However, it is also equally difficult to study in its full generality. This in result necessitated the use of some classical symmetry reductions to consider approximate models. Such models are ones which enjoy a certain symmetry, often in the context in cosmology \cite{Bondi:1947fta,Weinberg:1972kfs}, while another type of reduction is one in which gravity is considered in one dimension lower \cite{Gott:1982qg,Witten:1988hc,Witten:2007kt}. This is specifically interesting as, while the physics is different than that in 4-dimensions, this 3-dimensional theory is well understood and solvable \cite{Witten:1988hc}, often serving as a test case for candidate quantum theories of gravity, see for example \cite{Ashtekar:1989qd,Thiemann:1997ru,Carlip:1998uc,Witten:2007kt}. In all cases, a consistent quantum theory of gravity that is fully understood remains elusive due to grave technical and conceptual difficulties. 
\newline
Loop quantum gravity (LQG, see for example \cite{Rovelli:1997yv,Thiemann:2001gmi,Ashtekar:2004eh}) is an attempt to canonically quantise GR starting from its connection formulation with the fundamental variables taken to be the Ashtekar-Barbero connection and the densitised triads. The quantum states in LQG describe either 1 or 2-dimensional quantum gravitational excitations, for canonical-LQG or spinfoam-LQG respectively. One also has within LQG well defined notions of quantum geometric observables such as volume \cite{Rovelli:1994ge,Thiemann:1996au,Ashtekar:1997fb,Brunneman:2007as} and area \cite{Rovelli:1994ge,Ashtekar:1996eg}. Further, one has well defined quantum constraints which embody the Einstein field equations in the quantum theory \cite{Thiemann:1996aw,Thiemann:1996av,Thiemann:1997ru}. These quantum constraints come as a triplet: the Gau{\ss}, diffeomorphism (vector) and the Hamilton (scalar) constraints. The first generates gauge transformations, the second spatial diffeomorphisms. The dynamics are encoded in the quantum Hamilton constraint. 
\newline
Obtaining, understanding and interpreting the solutions to the dynamics of any quantum theory of gravity is a difficult task. Therefore unsurprisingly, the case is the same for LQG. Different regularisations for the quantum Hamilton constraint exist in the literature \cite{Thiemann:1996aw,Thiemann:1996av,Thiemann:1997rv,Thiemann:1997rt,Lewandowski:2014hza,Assanioussi:2015gka,Varadarajan:2021zrk,Varadarajan:2022dgg}. However, just as in the classical theory, general solutions are difficult to obtain leading to once again considering approximate models. Most notable outcomes is the development of loop quantum cosmology (LQC) which is the symmetry reduced FLRW cosmological spacetime quantised in LQG methods \cite{Bojowald:2001xe,Bojowald:2005epg,Ashtekar:2006wn,Ashtekar:2011ni}. Despite that, numerical methods are often employed, and so is the case in spinfoam-LQG as well \cite{Bahr:2016hwc,Dona:2019dkf,Han:2020npv,Dona:2022yyn,Han:2024rqb}.
\newline
Recently \cite{Sahlmann:2024pba}, novel numerical methods were used to solve the constraints of a gravity inspired quantum model. Specifically, a 3-dimensional $\Uone$ BF-theory was quantised using LQG methods and the neural network quantum state (NNQS) ansatz \cite{Carleo:2017nvk} was used to obtain solutions to the constraints of the canonical theory, making no use of classical symmetry reductions. It was demonstrated in \cite{Sahlmann:2024pba} that in principle one can apply such a variational ansatz to find solutions to a master constraint of this gravity inspired toy model considered with very high accuracy and efficiency.
\newline
The NNQS ansatz is a variational ansatz which allows one to harness the power of neural networks to arrive at ground states of the quantum Hamilton $\hat{H}$ of a quantum many-body system. This is done by finding the amplitudes in the many-body wave-function which minimise $\expect{\hat{H}}$ using a neural network. The NNQS was first considered in \cite{Carleo:2017nvk} where a restricted Boltzmann machine (RBM) \cite{10.1007/978-3-642-33275-3_2}, which is a two-layer generative stochastic network, was used to solve the Heisenberg model. Since then, it has showed remarkable success in several many-body physics problems, explored by different network architectures, and had its numerical efficiency compared to state of the art methods \cite{PhysRevB.100.125124,PhysRevB.97.035116,Choo_2020}. The key feature was the ability of this ansatz to parameterise the many-body wave-function with a number of parameters which is comparatively small in respect to the dimensions of the Hilbert space of the considered system.
\newline
During this recent study of applying NNQS to gravity-like models \cite{Sahlmann:2024pba}, the kinematical degrees of freedom were truncated by considering a fixed graph and only allowing a set of admissible representation labels\footnote{called charges and denoted as $m\in\Z$} $M := [-\jmax , \dots ,\jmax] \subseteq \Z$ for the holonomies in the models. This resulted in what was called a $\Uqone$ BF-theory where the $q$-deformation parameter was a root of unity. Even then, the Hilbert spaces, for large sets of admissible representation labels $M$ and/or arbitrarily large graphs, were rather large and growing exponentially with respect to the number of edges in the graph. The NNQS ansatz was nevertheless successful in solving such a model, at times with a number of parameters representing 1.57\% of the dimensions of the Hilbert space. Since the goal is to ultimately find approximate numerical solutions to the full 4d theory, this becomes a rather important, if not indispensable, feature of the numerical method of choice. We note that working on a fixed graph has also been advocated in algebraic LQG \cite{Giesel:2006uj,Giesel:2006uk,Giesel:2006um}. Of course, the graph employed there may be infinite or very large, much larger than the one we have to restrict to.  
\newline
In this work, we depart from the toy model considered in \cite{Sahlmann:2024pba} and inch closer to 4d gravity. We will consider 3d Euclidean gravity in the weak coupling limit of Smolin \cite{Smolin:1992wj} in which we will show is too a BF-theory but with a $\Uone^3$ gauge group. The purpose of this work is two-fold. First, it is to demonstrate that the NNQS ansatz can be utilised to solve different regularisations of the quantum Hamilton constraint. Second, we show that this ansatz and the computational framework developed in this work can model more complicated gauge groups. Both of these are building blocks for further work, as it is shown that 4d gravity in the weak coupling limit also has a gauge group of $\Uone^3$ \cite{Thiemann:2022all,Bakhoda:2022rut}. We will follow the implementation in \cite{Sahlmann:2024pba}, and define our quantum model on a fixed graph with similar truncation on the representation labels arriving at a 3-dimensional $\Uqone^3$ BF-theory. We will explore geometric observables of LQG, specifically the spatial volume operator in 3d Euclidean gravity as formulated in \cite{Thiemann:1997ru}. Further, we will consider two different regularisations for the quantum Hamilton constraint. The first being a master constraint constructed from the quantum analogs of the curvature and Gau{\ss} constraints and the second following the regularisation presented in \cite{Thiemann:1997ru}. We will solve both the constraints using the NNQS ansatz and we will compare the solution space of the first with the kernel of the second. The plan of the paper is then as follows:
\begin{enumerate}
    \item In Section \ref{sec:thePhysicalModel}, we present both classical and LQG quantised 3d Euclidean gravity in the weak coupling limit of Smolin. We also discuss the truncation of the kinematical degrees of freedom, the Thiemann regularisation of the quantum Hamilton constraint as well as the volume operator of 3d LQG.
    \item Next, the modelling of the now quantum theory and its constraints in the computational framework is provided in Section \ref{sec:computationalModel}.
    \item Following that, in Section \ref{sec:results}, we present the results obtained in this work in the following order:
    \begin{itemize}
        \item[$\bullet$] In Section \ref{sec:results_groundState}, the solution space of the master constraint of the quantum model is presented and discussed. We show that as the cutoff imposed on the labels of the holonomies is relaxed, the frustration between the Hamilton and Gau{\ss} constraints becomes alleviated and the behaviour of the continuum theory is obtained even for conservative cutoffs.
        \item[$\bullet$] Section \ref{sec:results_observables} concerns the observables of the theory where we will present the quantum fluctuations of the minimal loop holonomy operator. We show that as the cutoff is relaxed, the gauge invariant solutions obtained indeed become more flat. 
        \item[$\bullet$] In Section \ref{sec:results_volumeOperator} the quantum volume operator of 3d LQG is presented. We will examine its properties and behaviour in our quantum model and compare it with exact results expected from the literature when possible.
        \item[$\bullet$] The analysis of the kernel of the Thiemann regularised quantum Hamilton constraint is presented in Section \ref{sec:results_qhc} whereby we discuss the implementation of the constraint in the computational framework, the nature of the states near the kernel, and a detailed analysis comparing the obtained states to the solutions of the master constraint considered in Section \ref{sec:results_groundState}.
    \end{itemize}
    \item An appendix contains some observations about the scaling of Hilbert space dimension and about the scalar product of states picked randomly from a Hilbert space. 
\end{enumerate}

\section{Classical and quantum 3d Euclidean gravity in the weak coupling limit}
\label{sec:thePhysicalModel}
The starting point in this work is to consider the classical theory describing 3d Euclidean gravity, with no matter or cosmological constant contribution, and investigate the gauge group arising in Smolin's weak coupling limit \cite{Smolin:1992wj}. To obtain gravity as a BF-theory, we consider GR in the first-order formalism, in which GR is described as a theory of connections and vielbeins, rather than metrics. One such formulation is given by the Palatini action. For a spacetime of $n$-dimensions $\spacetime = \R\times\Sigma^{(n-1)}$, one can choose the fundamental variables to be an orthonormal frame of 1-forms $e^I$ and a metric compatible connection $\omega^I{}_J$, where we use $I, J, \dots$ to denote the internal indices. In ($n-1$) + 1-dimensions, the Palatini action is expressed as \cite{Montesinos:2019bkc}
\begin{equation}
    S[e, \omega] = \frac{1}{\kappa}\int_{\spacetime} \star \left(e^I \wedge e^J\right) \wedge F_{IJ}(\omega),
\end{equation}
where $F_{IJ}(\omega) = \extd \omega^I{}_J + \omega^I{}_{K} \wedge \omega^K{}_{J}$ is the curvature tensor of the connection, $\kappa$ is a constant related to Newton's gravitational constant $G_N$ and the Hodge dual map is given by \cite{Montesinos:2019bkc}
\begin{equation}
    \star(e_{I_1} \wedge \dots \wedge e_{I_k}) := \frac{1}{(n-k)!} \epsilon_{I_1 \dots I_k I_{k+1} \dots I_n} e^{I_{k+1}} \wedge \dots \wedge e^{I_n}.
\end{equation}
For a metric $\eta_{IJ} := \mathrm{diag}(\sigma, 1 \dots , 1)$ with $\sigma = +1$, the frame rotation group is the SO($n$) group while for $\sigma = -1$, it is the Lorentz group SO($n - 1$, $1$). Here, $\epsilon$ is the totally antisymmetric tensor. The equations of motion arising from this action are equivalent\footnote{for non-degenerate orthonormal frames} to the Einstein field equations. Now, in 3-dimensions the Palatini action takes the simple form
\begin{equation}
\label{eq:palatini2+1}
    S[e, \omega] = \frac{1}{\kappa} \int_\spacetime e^{K} \wedge F_{K} = \frac{1}{\kappa} \int_\spacetime \left(e^I \wedge \extd\omega_I + \frac{1}{2}\epsilon_I{}^{JK} e^I\wedge\omega_J \wedge\omega_K \right). 
\end{equation}
The canonical analysis on this action shows that the phase space is subject to the two first-class constrains which are written as \cite{Corichi:2015nea,Ashtekar:1989qc}
\begin{equation}
    F^I = 0 \quad , \quad \extd^{(\omega)} e^I = 0,
    \label{eq:constraints}
\end{equation}
where $\extd^{(\omega)}$ is the covariant exterior derivative with respect to $\omega$. The first of the two equations shown in  \eqref{eq:constraints} is the \textit{curvature constraint} enforcing flatness of the connection and the second is the \textit{Gau{\ss} constraint} generating gauge transformations. In our work, we will consider Euclidean gravity, hence we take the Riemannian signature of the metric. Thus, the configuration space is the space of SO(3)-valued connections and the phase space is the cotangent bundle over it. The Gau{\ss} constraint generates SO(3) transformations.
\newline
It is clear that the Palatini action in 3-dimensions shown in equation \eqref{eq:palatini2+1} is of BF-type \cite{Celada:2016jdt}, and hence a topological theory, whereby the Lie group is SO(3) and the role of $B$ is played by the orthonormal frame of 1-forms $e^I$. In the absence of a cosmological constant, the solutions are locally flat (homogeneously curved otherwise) and the theory has no local degrees of freedom. In the weak coupling limit of Smolin \cite{Smolin:1992wj}, the action \eqref{eq:palatini2+1} reduces to
\begin{equation}
\label{eq:palatiniWeakCoupling}
    S[e, \omega] = \int_\spacetime e^{I} \wedge \extd \Tilde{\omega}_{I},
\end{equation}
where $\Tilde{\omega} = \omega / \kappa$. What is observed is that one does not have any quadratic terms of the connection arising. This is also reflected in the constraints. In this limit, the constraints pulled back to $\Sigma$ are expressed as
\begin{equation}
F^{I}_{ab} := 2\partial_{[a}\Tilde{\omega}^{I}_{b]} = 0 \quad , \quad \partial_{a} e^{a}_{I} = 0
    \label{eq:constraintsWeakCoupling}
\end{equation}
where $a, b, \dots = 1, 2$ denote spatial indices. To investigate the type of gauge transformations generated by the Gau{\ss} constraint shown on the right in \eqref{eq:constraintsWeakCoupling}, let $\Lambda$ be a smearing field and define on the phase space the functional
\begin{equation}
    G(\Lambda) = \int_{\Sigma} d^2x \Lambda_{I} \partial_b e^I_c \varepsilon^{bc}.
\end{equation}
Then, one can compute the Poisson bracket of $G(\Lambda)$ with the connection to see that
\begin{align}
    \lbrace \omega^{I}_a(x), G(\Lambda)\rbrace  
    & = \int_{\Sigma} d^2y \partial_b^{(y)} \lbrace \omega^{I}_{a}(x),  e^{J}_c(y) \rbrace \varepsilon^{bc} \Lambda_J (y)\\
    & =- \partial_a \Lambda_I (x).
\end{align}
So $\omega$ transforms like a $\Uone^3$ connection under the transformations generated by $G(\Lambda)$. Note also  
$\lbrace e^{I}(x), G(\Lambda)\rbrace = 0$ as expected for an Abelian gauge theory. Thus, in this limit, the theory is $\Uone^3$ BF-theory on a 3-dimensional manifold $M^{(3)} = \R \times M^{(2)}$. In the following, we will take $M^{(2)} = \R^2$. In what follows, we canonically quantise this model using LQG methods.

\subsection{Quantum $\Uone^3$ BF-theory}
\label{sec:lqg}
LQG uses the formulation of GR in which the (kinematical) phase space is parameterised by the Ashtekar-Barbero connection $A^I_a$ and the densitised triads $E^I_a$ \cite{Ashtekar:2004eh}. This means that the physical phase space of GR is embedded in that of SU(2) Yang-Mills theory and facilitates the use of familiar tools from gauge theory in LQG. Holonomies of $A^I_a$ and fluxes of $E^I_a$ are quantized and lead to a closed algebra \cite{Ashtekar:2004eh}.
\newline
To outline this process, let $c: [0, 1] \rightarrow \Sigma$ denote a $C^k$ semi-analytic path embedded in $\Sigma$ ($k\gg 1$). A holonomy is a $G$-valued element here denoted $h_c (A)$ where $A$ is a $\mathrm{Lie}(G)$-valued connection 1-form, the Ashtekar-Barbero connection in the case of LQG. They are nothing but the path ordered exponential integral of the connection along paths $c$. Important functions of holonomies are their matrix elements in irreducible representations of $G$ which are hereby denoted by $\pi_j$ with a $G$ representation label $j$. For brevity, we will denote by $h^j_c (A) := \pi_j (h_c (A))$. Smearing the densitised triads over 2-surfaces $S$, the pair of variables then can be written as \cite{Ashtekar:2004eh}
\begin{equation}
\label{eq:HF}
    h_c[A] = \mathcal{P}\exp\left(-\int_c A\right) \quad , \quad E_{S,i}[E] = \int_S E^{a}_i dS_a.
\end{equation}
Holonomies can be seen to transform under gauge transformations only at their endpoints. That is, for some transformation $U(x) \in G$,
\begin{equation}
    h^j_c(A) = U^{-1}(c(1))\; h^j_c(A) \; U(c(0)).
\end{equation}
The pair \eqref{eq:HF} has Poisson brackets which yields sums of products of functions of holonomies \cite{Ashtekar:2004eh}. This holonomy-flux algebra can be represented on a Hilbert space $\hilb = L^2 (\mathcal{A}, \extd \mu)$ where $\mathcal{A}$ is the space of distributional connections and a measure $\mu$ \cite{Ashtekar:2004eh,Thiemann:2001gmi}. A dense subset of this space is given by cylindrical functions $\Psi \in \mathrm{Cyl}$ which are complex valued functions which depend on $A$ through a finite number of holonomies
\begin{equation}
    \Psi [A] = \psi (h_{e_1}[A] , \dots , h_{e_n}[A]).
\end{equation}
where $\psi : G^n \rightarrow \mathbb{C}$. Here, the collections of paths $e_1 , \dots , e_n$ are denoted the name edges, and they only meet at their end/start points which are denoted the name vertices $v$. Together, they are said to constitute a graph $\gamma \hookrightarrow \Sigma$. 
\newline
To construct gauge invariant states, one chooses a $G$-invariant tensor, the intertwiner $\iota_v$, for every vertex $v$, which intertwines the matrix elements of the holonomies of the incoming and outgoing edges at $v$ in an appropriate manner \cite{Thiemann:2021hpa}. One can then define spin-network functions (SNFs) on $\gamma$ which are functions composed of products of the $h_e^j$ for every $e \in E(\gamma)$ contracted with suitable intertwiners for every vertex \cite{Thiemann:2021hpa}. 
\newline
If we now consider the specific case $G = \Uone^3$ then the representation labels are charge vectors $\Vec{m} \in \Z^3$. The intertwiner spaces are one- or zero-dimensional, depending on whether the incoming charges match the outgoing ones \cite{Thiemann:2021hpa,Bakhoda:2022rut}. The state is gauge invariant iff  
\begin{equation}
    \sum_{e \text{ in at } v} \vec{m}_e=  \sum_{e \text{ out at } v} \vec{m}_e. 
\end{equation}
for all vertices $v$. In this case we call the state a charge-network function (CNF) on $\gamma$ and denote it by $\cnf_{\gamma, \lbrace \Vec{m} \rbrace}$. Explicitly, 
\begin{equation}
    \cnf_{\gamma, \lbrace \Vec{m} \rbrace}[A] = \prod_{e \in E(\gamma)} h^{\Vec{m}_e}_e [A].
\end{equation}
The kinematical Hilbert space $\hilb_{\mathrm{AL}} = L^2(\mathcal{A}, \extd \mu_{\mathrm{AL}})$ is then taken to be spanned by the CNFs along with the Ashtekar-Lewandowski measure \cite{Thiemann:2021hpa,Bakhoda:2022rut}. On such a space, the flux operator acts as a derivation with respect to a holonomy \cite{Bakhoda:2022rut}. Further, the holonomy operator acts as a multiplication operator, merely changing the label on the corresponding edge it acts upon \cite{Bakhoda:2022rut}. Precisely, this can be written as
\begin{equation}
\label{eq:h_action}
    \hat{h}_{e_k}^{\Vec{m}_0} \cnf_{\gamma , \lbrace \Vec{m} \rbrace} = \cnf_{\gamma, \lbrace \Vec{m}' \rbrace},
\end{equation}
where now 
\begin{equation}
\label{eq:h_actionn}
\lbrace \Vec{m}' \rbrace = \lbrace \Vec{m}_{e_1}, \Vec{m}_{e_2}, \dots , \dots \Vec{m}_{e_k} + \Vec{m}_0 , \dots , \dots , \Vec{m}_{|E(\gamma)|} \rbrace,
\end{equation}
if $e_k \in E(\gamma)$ otherwise it modifies the graph by adding an edge $e_k$. 
\newline
The constraints of this resulting quantum theory are composed of the quantum analogs of the curvature and Gau{\ss} constraints. We will discuss them below, in section \ref{sec:constraints}.

\subsection{Quantum $\Uqone^3$ BF-theory}
The last step of quantising this theory in our work is one necessitated due to computational resources: truncating the number of kinematical degrees of freedom. This comes in two forms, just as in \cite{Sahlmann:2024pba}. First, we restrict our model to only one fixed graph $\gamma$. Further, we employ a cutoff on the irreducible representations of $\Uone^3$. We denote the set of admissible charges by
\begin{equation}
    M:= [-\jmax, \dots, \jmax] \subseteq \Z,
\end{equation}
where $\jmax \in \Z$. As such, we have a total of $2\jmax + 1$ possible choices for every component and a total of $(2\jmax + 1)^3$ possible values for a given charge vector. To do this in a consistent and elegant fashion, we interpret these as labels for irreducible representations of the quantum group $\Uqone^3$. This entails a deformation of the dual group of irreducible representations and hence also of the holonomy operators. The tensor product of irreducible representations is now 
\begin{equation}
\label{eq:pbc}
    m_i \otimes n_i := (m_i + n_i + \jmax) \mod (2\jmax + 1) - \jmax, \qquad i=1,2,3.
\end{equation}
This also dictates a deformation of the action \eqref{eq:h_action}, \eqref{eq:h_actionn} of the holonomies on quantum states. We should note, however that this also means the gauge group is now $\Uqone^3 \simeq \Z^3_{2\jmax + 1}$ \cite{2012arXiv1209.1135G,Sahlmann:2024pba}, in the following sense: While $\Uone^3$ still acts on the resulting Hilbert space, and we will be implementing the constraints of the continuum theory, the action of the holonomy operators is not gauge covariant under $\Uone^3$, but only under $\Uqone^3$. 
\newline
Let us briefly dwell on a possible physical meaning of the cutoff. The $q$-deformation parameter is expressed as \cite{2012arXiv1209.1135G}
\begin{equation}
    \label{eq:qValue}
    q = t^2 \qquad , \qquad t = \exp\left({\frac{i\pi}{2\jmax + 1}}\right).
\end{equation}
This could be compared to the Turaev-Viro model \cite{Turaev:1992hq,Barrett:1993ab,Dittrich:2018dvs} in which a $q$-deformation of SU(2) with
\begin{equation}
\label{eq:qValTV}
    q = t^2 \qquad , \qquad t = \exp\left(\frac{i\pi}{k + 1}\right).
\end{equation}
encodes a cosmological constant \cite{Dittrich:2018dvs}
\begin{equation}
\label{eq:cosmConst}
    \Lambda = \frac{1}{G_N^2 \hbar^2 k^2}.
\end{equation}
$k$ is the level of the Chern-Simons theory describing 3d gravity. By comparing \eqref{eq:qValue} and \eqref{eq:qValTV}, $k$ in our work corresponds to $2\jmax$. This in turn implies that an effective cosmological constant would be 
\begin{equation}
    \Lambda = \frac{1}{4G_N^2 \hbar^2 \jmax^2}.
\end{equation}
in a given cutoff $\jmax$. The interpretation drawn is then that for larger $\jmax$, one approaches a setting where $\Lambda$ is small. Of course, this has to be taken with a lot of caution because the weak coupling limit employed here might have very little to do with the original theory. 
\newline
Lastly, one surprising consequence of this is that the curvature and Gau{\ss} constraints will not commute in the quantum theory. Consequently, for finite $\jmax$ there are no exact joint solutions of all the constraints. This frustration will however be shown to be alleviated for large $\jmax$. We will, in the following refer to this gravitational model as quantum $\Uqone^3$ BF-theory.

\subsection{Geometric operators}
As previously mentioned, LQG comes with well-defined quantum geometric observables. Therefore, in this quantised model one has for example a well-defined notion of quantum volume. For a compact region $B \subset \Sigma$, the volume operator is expressed as \cite{Thiemann:1997ru}
\begin{equation}
    \hat{V}(B) := \sum_{v \in V(\gamma) \cap B} \hat{V}_{v} \,,\qquad \hat{V}_v := \sqrt{\sum_I\left( \sum_{e, e' \text{ at } v} \sign(e, e') \epsilon_{IJK} X^{J}_{e} X^{K}_{e'}\right)^2},
\end{equation}
where in the $\Uone^3$-limit, the $X^{I}_e$ are the three invariant vector fields on $\Uone^3$, acting on holonomies along $e$, and $\sign(e, e')$ is the orientation of the tangents of the ordered pair $(e, e')$ relative to the orientation of $\Sigma$. 
\newline
In the 4d case, an analog of this operator would vanish at gauge invariant vertices of valence 3 or less while in 3-dimensions, this the case for valence 2 only, so long as there are at least a pair of edges incident at the vertex with linearly independent tangents. In this work, we will explore the implementation of this operator in our computational framework and examine its properties as given in the literature. This operator also plays a role for the Thiemann regularised quantum Hamilton constraint which we also consider in this work.

\subsection{Constraint operators}
\label{sec:constraints}
The classical theory considered so far is subject to the curvature and the Gau{\ss} constraints \eqref{eq:constraints}. The Gau{\ss} constraints imposes charge vector conservation at every vertex of the graph. The curvature constraint ensures flatness of holonomies. 
In one of the two approaches pursued in this work, we impose both constraints in the form of master constraints \cite{Thiemann:2003zv,Dittrich:2004bn,Thiemann:2005zg}. For a fixed graph $\gamma$ 
\begin{equation}
\label{eq:F_master}
     \hat{F}_\gamma  = \sum_{\alpha \in L(\gamma)} \tr \left[\left(\hat{h}_{\alpha} - \one\right)\left(\hat{h}_{\alpha}^{\dagger} - \one\right)\right]
\end{equation}
where $L(\gamma)$ is a set of minimal loops in $\gamma$, and 
\begin{equation}
\label{eq:gaussoperator}
    \hat{G}\rvert_\gamma = \sum_{v\in V(\gamma)} \sum_{i=1}^3 (\hat{E}_{S(v),i})^2 
\end{equation}
where $S(v)$ is a small sphere around the vertex $v$. In principle, we would then demand 
\begin{equation}
    \label{eq:solutiona}
    \bra{\Psi_\gamma}\hat{F}_\gamma + \hat{G}\rvert_\gamma\ket{\Psi_\gamma} =0 
\end{equation}
for a physical state $\Psi_\gamma$. 
Just as in \cite{Sahlmann:2024pba}, the solutions of the flatness constraint are not normalisable in $\hilb_{\mathrm{AL}}$ but instead form a different measure $\mu_\text{flat}$ which is the $\delta$-measure on flat connections \cite{Ashtekar:1994mh,Sahlmann:2011xu,Dittrich:2014wpa,Drobinski:2017kfm}. It is not absolutely continuous with respect to the Ashtekar-Lewandowski measure but we also nevertheless choose to work with states in $\hilb_\text{AL}$, which can approximate solutions. For details of how these constraints are implemented in the computational model, see Section \ref{sec:implementation}.
\newline
The constraints \eqref{eq:constraints} look structurally different from those of 4d gravity \cite{Yoon:2019kiv}. Nevertheless, it was shown that one can bring them into the same form \cite{Thiemann:1997ru}. This is very interesting, because it maximizes the analogy of this model to 4d gravity. In  this form, the constraints read \cite{Thiemann:1997ru}
\begin{equation}
    G_{I} := \mathcal{D}_{a}E^{a}_{I} \quad , \quad V_{a} := F^{I}_{ab} E^{b}_{I} \quad , \quad H := \frac{1}{2\sqrt{\det q}} \epsilon_{IJK} F^{I}_{ab} E^{a}_{J} E^{b}_{K}
    \label{eq:newConstraints}
\end{equation}
where $\mathcal{D}_a$ denotes the covariant derivative with respect to the connection, $q$ is spatial the metric and $E_I^a$ are modified densitised triads \cite{Thiemann:1997ru}. The Gau{\ss} constraint remains unchanged. However, the curvature constraint is now replaced by the diffeomorphism (vector) constraint $V_a$ and the Hamilton (scalar) constraint $H$. The field theories governed by the constraints \eqref{eq:constraints} shown previously and these three constraints shown above are classically equivalent. However, it is argued that upon quantisations, the two quantum theories are different, having little overlap in their solution spaces \cite{Thiemann:1997ru}. 
\newline
The exact procedure of deriving the quantum Hamilton constraint, is too complex to be presented here, and consequently, we only state the result. We partially follow the prescription proposed in \cite{Thiemann:1997ru}: The classical constraint is regularized in terms of holonomies, fluxes, and their Poisson brackets, all associated to a triangulation $T$. 
To turn this regularization into an operator, one adapts the triangulation to a graph, $T\equiv T(\gamma)$, replaces holonomies and fluxes by their operator counterparts and takes a certain limit:
\begin{align}
    \hat{H}_{T(\gamma)}(N) f_\gamma & := \lim_{\epsilon \rightarrow 0} \hat{H}_{T, \epsilon}(N) f_\gamma \\
    & = \frac{2}{\hbar^2}\sum_{\Delta, \Delta' \in T, v} \epsilon^{ij}\epsilon^{kl}N(v) \tr(\hat{h}_{\alpha_{ij}(\Delta')} \hat{h}_{s_{k}(\Delta)} [\hat{h}_{s_{k}(\Delta)}^{-1}, \sqrt{\hat{V}_v}] \hat{h}_{s_{l}(\Delta)} [\hat{h}_{s_{l}(\Delta)}^{-1}, \sqrt{\hat{V}_v}]) f_\gamma.
    \label{eq:trc}
\end{align}
Here, $\Delta, \Delta' \in T$ are solid triangles sharing a common basepoint $v(\Delta) = v(\Delta') =: v$, $\alpha_{ij}(\Delta')$ is a loop created using a specific prescription along the edges labeled by $i, j$ of $\Delta'$ which meet at $v$, $s_k, s_l$ are edges in $\Delta$ meeting at $v$, $N$ is a smearing function and $f_\gamma$ are functions on the graph \cite{Thiemann:1997ru}. 
\newline
Solutions to the constraint are by definition diffeomorphism invariant states $(\Psi\rvert$ such that 
\begin{equation}
\label{eq:solution}
(\Psi\rvert \hat{H}_{T(\gamma)}(N)\ket{f_\gamma}=0 \text{ for all } \gamma, f_\gamma \in \hilb_\gamma, N. 
\end{equation}
For the purpose of our work, this prescription poses a problem, since the operator defined in this way is graph-changing, whereas we would like to work on a fixed graph. In \cite{Thiemann:1997ru}, the loop $\alpha_{ij}$ is formed by taking two edges $i, j$ of a solid triangle $\Delta \in T$ and creating a new edge connecting them. This edge is usually created halfway through the edges $i, j$. For our purposes, it is natural to change this prescription, by closing the holonomy along a loop that is already part of $\gamma$. Thus in the following, $\alpha_{ij}$ denotes a minimal loop in $\gamma$, with basepoint $v$ that starts with edge $e_i$ and ends with edge $e_j$. We will still refer to this constraint as the Thiemann regularised constraint (TRC for short) owing to the work done in \cite{Thiemann:1997ru}. We note that this operator need not be self-adjoint, nor non-negative. We will also see this in the example implemented below. 
\newline
Since we will use a graph non-changing operator, we can somewhat simplify the condition \eqref{eq:solution}. Solutions $(\Psi\rvert$ must be the diffeomorphism average of some cylindrical function, $(\Psi\rvert=(\Psi_{\gamma_0}\rvert$. But then, due to the properties of the group averaging and the kinematical inner product (see for example \cite{Ashtekar:2004eh}), \eqref{eq:solution} is already fulfilled if $\gamma_0$ is not diffeomorphic to $\gamma$\footnote{Strictly speaking, although our regularization of $H$ can still annihilate edges of $\gamma$, although it uses loops that are part of the graph: It could so happen that it cancels all the charges on that edge. Thus, strictly speaking, we also have to consider the case where $\gamma_0$ is diffeomorphic to a subgraph of $\gamma$. Since this annihilation of edges only happens in highly non-generic situation, we will not consider it in the following.}. Then, again due to the properties of the group averaging \cite{Ashtekar:2004eh}, \eqref{eq:solution} is equivalent to  
\begin{equation}
\label{eq:solutionnnn}
\bra{\Psi'_\gamma} \hat{H}_{T(\gamma)}(N)\ket{f_\gamma}=0 \text{ for all } f_\gamma \in \hilb_\gamma, N, 
\end{equation}
where we have introduced a modified cylindrical function $\Psi'_\gamma$ which is chosen such that it gives the same group averaged state, 
$(\Psi_\gamma|=(\Psi'_\gamma|$
and that it is invariant under the action of the graph symmetries \cite{Ashtekar:2004eh} of $\gamma$. 
This in turn is equivalent to 
\begin{align}
\bra{\hat{H}^\dagger_{T(\gamma)}(N)\Psi'_\gamma} \ket{f_\gamma}&=0 \text{ for all } f_\gamma \in \hilb_\gamma, N\\
\Longleftrightarrow \qquad 
\hat{H}^\dagger_{T(\gamma)}(N)\Psi'_\gamma & = 0 \text{ for all } N\\
 \Longleftrightarrow \qquad
\bra{\Psi'_\gamma} \hat{H}_{T(\gamma)}(N)\hat{H}^\dagger_{T(\gamma)}(N)\ket{\Psi'_\gamma}&=0 \text{ for all } N. \label{eq:strong_solution}
\end{align}
Now \eqref{eq:strong_solution} finally has the right form for the numerical methods that we are going to deploy. Alas, the complexity of $\hat{H}_{T(\gamma)}$ is such that its modulus squared is beyond the capabilities of the code that we are going to deploy. We will instead look for solutions of 
\begin{equation}
    \bra{\Psi_\gamma} \hat{H}_{T(\gamma)}(N) + \hat{H}^\dagger_{T(\gamma)}(N)\ket{\Psi_\gamma}=0, 
\end{equation}
for fixed $N$. Thus we will only implement a necessary condition for $\Psi_\gamma$ to be a solution, not a sufficient one. 
\newline
To summarize, we have quantized 3d Euclidean gravity in the weak coupling limit using LQG methods and the kinematical degrees of freedom were truncated. The constraints are expressed in two different ways: On the one hand as master constraints \eqref{eq:solutiona}. On the other hand, we will implement the Gau{\ss} constraint as a master constraint \eqref{eq:gaussoperator} the same way, but treat diffeomorphism and Hamilton constraints \eqref{eq:newConstraints} in a way that resembles loop quantum gravity in 4d, see \eqref{eq:trc}, \eqref{eq:solution}.

\section{The computational model}
\label{sec:computationalModel}
This section covers the two topics of (i) implementing the physical model in a computational framework and (ii) the network architecture used in this work. In (i), we will present the expression of the constraints of the quantum $\Uqone^3$ BF-theory which are to be implemented. Furthermore, the exact setup to accommodate for the different gauge group is discussed.

\subsection{The implementation of the physical model}
\label{sec:implementation}
\begin{figure}[h]
    \centering
    \includegraphics[width = 0.85\linewidth]{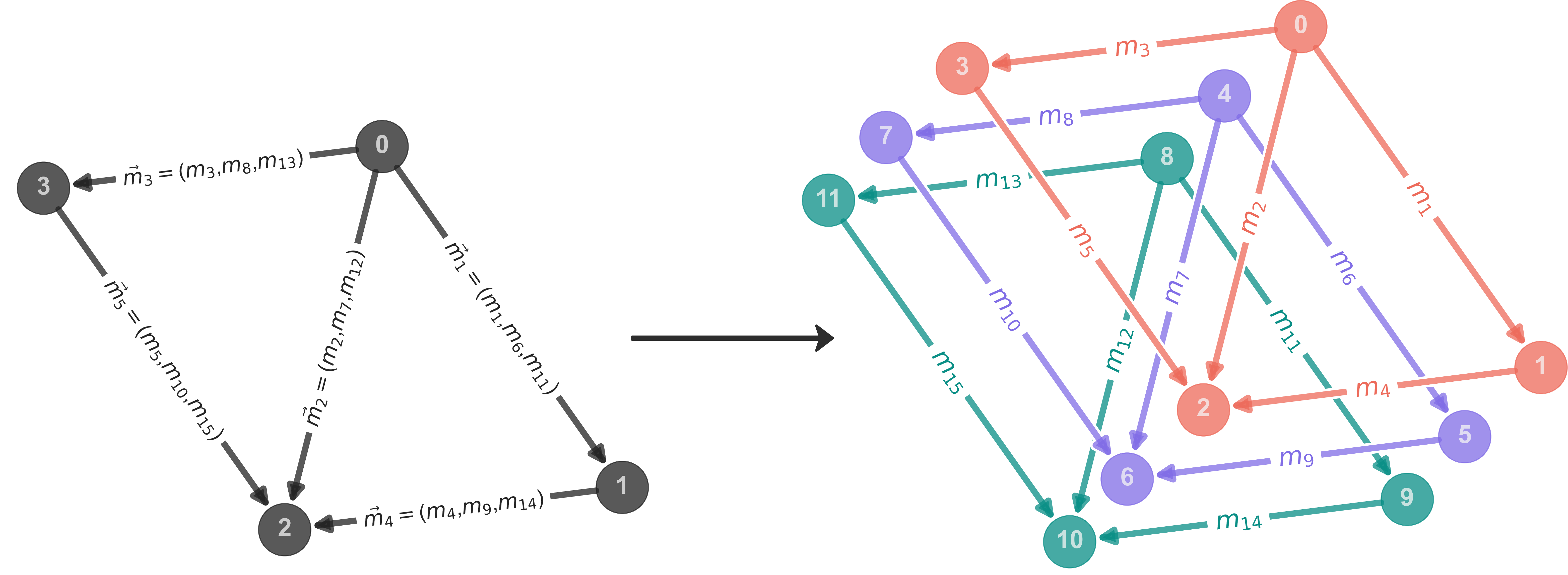}
    \caption{On the left, the graph $\gamma$ on which the analytical model would be based upon. On the right, the equivalent implementation on the computational framework. The analytical graph $\gamma$ is implemented as three different independent computational graphs, all identical to $\gamma$. Every computational graph carries the set of charge numbers corresponding to certain components of the charge vectors of $\gamma$.}
    \label{fig:threeGraphs}
\end{figure}

In this work, we will adopt both the terminology and the implementation provided in \cite{Sahlmann:2024pba}. Our models are to be defined in the computational framework on graphs denoted by $\Tilde{\gamma}$ which are the dual to an oriented graph $\gamma$. Here, duality means that every edge and vertex in $\gamma$ is replaced by a dual vertex $\Tilde{v}$ and a dual edge $\Tilde{e}$ in $\Tilde{\gamma}$ respectively. We will consider non-trivial $N$-L graphs where the number of minimal loops\footnote{the set of smallest length cycles in the cycle basis of the cycle space of the undirected version of $\gamma$} $N = |L(\gamma)| \geq 2$\footnote{For $N = 1$, the Gau{\ss} constraint is already perfectly imposed}. The reason for the implementation on the dual graph is purely driven by computational feasibility: as the edges in $\gamma$ carry the charge vectors, it is easier to model the Hilbert space as the tensor product of the Hilbert spaces of the dual vertices. Further, minimal loops are taken to have the same incrementing orientation discussed in \cite{Sahlmann:2024pba} and as such, the orientation of the loop need not coincide with that of the graph.
\newline
Now, one has at hand a many-body system where the role of a particle is played by a dual vertex. Every dual vertex has a finite set of $(2\jmax + 1)^3$ degrees of freedom for the charge vector associated to it. As such, for a dual graph of $K$ dual vertices, the dimensions of the Hilbert space over $\gamma$ would be $\dim\hilb_{\gamma} = (2\jmax + 1)^{3K}$, thrice as large compared to what is considered in the $\Uqone$ model \cite{Sahlmann:2024pba}. Here, we have imposed the same cutoff conditions as in \cite{Sahlmann:2024pba} as shown also in equation \eqref{eq:pbc} above where each component of the charge vector is independently restricted within a set of allowed charges $M = [-\jmax, \dots , \jmax] \subseteq \Z$. 
\newline
The next point of order is to outline the implementation of now the three dimensional gauge group. Consider once again the simple $\Uqone$ BF-theory in \cite{Sahlmann:2024pba} where a dual graph composed of 5 dual vertices is chosen. Then, the basis states would be a 5-fold tensor product of basis states in the Hilbert spaces $\hilb_{\Tilde{v}}$ living on the dual vertices. Since the basis states of $\hilb_{\Tilde{v}}$ are labeled by the charge numbers, a basis state in the total Hilbert space over the entire graph was given by a quintuplet
\begin{equation}
    (m_1, m_2, m_3, m_4, m_5) := \ket{m_1} \otimes \ket{m_2} \otimes \ket{m_3} \otimes \ket{m_4} \otimes \ket{m_5},
\end{equation}
where $m_k$ denotes the charge on the dual vertex $k$ and $m_k \in [-\jmax, \dots, \jmax] \subseteq \Z$. In the present case, one has for every dual vertex a charge vector. As such, a basis state in the $\Uqone^3$ model considered here over the same graph would be identified with five charge vectors, or equivalently 15 charge numbers
\begin{equation}
\label{eq:basisStateUq3}
    (\underbrace{m_1^1, m_2^1, m_3^1}_{\Vec{m}_1}, \underbrace{m_1^2, m_2^2, m_3^2}_{\Vec{m}_2}, \underbrace{m_1^3, m_2^3, m_3^3}_{\Vec{m}_3}, \underbrace{m_1^4, m_2^4, m_3^4}_{\Vec{m}_4}, \underbrace{m_1^5, m_2^5, m_3^5}_{\Vec{m}_5}).
\end{equation}
where $m_k^l$ denotes the $k^{\text{th}}$ component of the charge vector $\Vec{m}_l$. Thus, one needs to label the basis states of the dual vertices in the computational framework with the charge vectors $\Vec{m} \in \Z^3$. However, in our computational framework, this is not feasible. Furthermore, since the charge numbers for every component of a given charge vector can be the same, one cannot uniquely map the charge numbers of the components of a given charge vector to another basis in an isomorphic manner that allows for distinguishing between different components. The workaround implemented in this work is that one considers three different copies of the same graph, with every copy of the dual vertex in every different copy of the graph carrying one component of the charge vector. 
To illustrate this, Figure \ref{fig:threeGraphs} shows an example of this decomposition of the charge vectors over three graphs in the computational framework.
\newline
In Figure \ref{fig:threeGraphs}, the graph on which the analytical model is defined can be seen on the left in black, where each edge carries a charge vector composed of three charge values. The equivalent computational implementation of such a graph is shown on the right where we have three copies of $\gamma$ where now each edge in every copy carries one of the charge numbers in the charge vectors on $\gamma$. Every copy of the graph is then implemented independently, with its own dual graph, in the computational framework. The operators in the constraints then act on the appropriate graphs as they should.
\newline
We now move to discuss the implementations of the constraints of the $\Uqone^3$ BF-theory. We recall that the master constraint $\hat{C}$ considered in \cite{Sahlmann:2024pba} was composed of a the Gau{\ss} constraint $\hat{G}$ and curvature constraint $\hat{F}$. To further facilitate the computational implementation, the curvature constraint imposed flatness on the minimal loops as one can faithfully decompose any loop in the graph in terms of the minimal loops.
\newline
In the current work, the curvature constraint \eqref{eq:F_master}
also imposes flatness over the minimal loop holonomies. Let ${}^{(1)}$ denote the operators in the $\Uqone$ theory and introduce the notation
\begin{equation}
    \hat{A}^{(1)_m} := \overbrace{\underbrace{\one \otimes \one \otimes \dots \otimes \one \otimes}_{(m - 1)-\text{ fold}} A^{(1)} \otimes \one \otimes \dots}^{n-\text{ fold}} ,
\end{equation}
to denote that the operator $\hat{A}^{(1)}$ sits in the $m$\textsuperscript{th} position in the $n$ fold tensor product shown above. In our work, $n = 3$. The curvature constraint  \eqref{eq:F_master} can then be expressed as
\begin{align}
    \hat{F} & = \sum_{\alpha \in L(\gamma)} \tr \left[\left(\hat{h}_{\alpha} - \one\right)\left(\hat{h}_{\alpha}^{\dagger} - \one\right)\right] \\
    & = \sum_{\alpha \in L(\gamma)}\sum_{k = 1}^{3} \left(\hat{h}_{\alpha}^{(1)_k} - \one\right) \left(\hat{h}_{\alpha}^{(1)_k^\dagger} - \one\right) \\
& = \sum_{\alpha \in L(\gamma)}\sum_{k = 1}^{3} \left(2\one - \hat{h}_{\alpha}^{(1)_k} - \hat{h}_{\alpha}^{(1)_k^\dagger}\right) \\
& = \sum_{\alpha \in L(\gamma)}\left(6\one - \tr \hat{h}_{\alpha} - \tr \hat{h}_{\alpha}^{\dagger}\right) \\
& = 6|L(\gamma)|\one - \sum_{\alpha \in L(\gamma)} \tr\left[\hat{h}_{\alpha} + \hat{h}_{\alpha}^{\dagger}\right]
\end{align}
where we denote by
\begin{equation}
    \tr \hat{h}_{\alpha} := \sum_{k = 1}^{3}\hat{h}_{\alpha}^{(1)_k}.
\end{equation}
For brevity, we will often refer to $\tr \hat{h}_\alpha$ as the minimal loop holonomy operator despite it actually being the trace of it. Recall that the $\Uqone$ minimal loop holonomy operators are of the form \cite{Sahlmann:2024pba}
\begin{equation}
    \hat{h}_{\alpha}^{(1)} = \prod_{\Tilde{v} \in \Tilde{V}(\alpha)} \overset{\leftrightarrow}{\hat{h}^{(1)}_{\Tilde{v}}}. 
\end{equation}
Here, the $\leftrightarrow$ indicates that one uses either the holonomy operator or its adjoint depending on the orientation of the edges of minimal loop relative to the underlying graph \cite{Sahlmann:2024pba}. These basic $\Uqone$ holonomy operators $\hat{h}^{(1)}_{\Tilde{v}}$ are defined as in \cite{Sahlmann:2024pba}, acting as raising and lowering operators for the charge numbers while imposing the same periodic boundary conditions discussed therein and mentioned in equation \eqref{eq:pbc} above.
\newline
Since the vector addition/subtraction of charge vectors is defined element-wise, the Gau{\ss} constraint is imposed on every copy of the graph independently as well. Thus, we once again have three of the Gau{\ss} constraint as implemented in \cite{Sahlmann:2024pba} which gives the expression for the $\Uqone^3$ Gau{\ss} constraint \eqref{eq:gaussoperator} 
\begin{equation}
    \hat{G} = \hat{G}^{(1)_1} + \hat{G}^{(1)_2} + \hat{G}^{(1)_3},
\end{equation}
where by $\hat{G}^{(1)_1}$ we mean that it is the $\Uqone$ Gau{\ss} constraint imposed on the first graph. We recall that \cite{Sahlmann:2024pba}
\begin{equation}
    \hat{G}^{(1)} = \sum_{v \in V(\gamma)} \left(\sum_{e \in E_i(\gamma)} \hat{N}_{e}^{(1)} - \sum_{e' \in E_o(\gamma)} \hat{N}_{e'}^{(1)}\right)^2.
\end{equation}
where $E_i (v), E_o (v)$ indicates the set of edges incident at and outgoing from a vertex $v$ respectively. Here, the $\hat{N}^{(1)}$ operator is defined as in \cite{Sahlmann:2024pba}, which is nothing but a diagonal $\dim\hilb_{\gamma} \times \dim\hilb_{\gamma}$ matrix with the diagonal being $\mathrm{diag}(\jmax, \dots, -\jmax)$ which upon acting on an edge just gives the charge that the edge carries. 
\newline
Now, the master constraint considered in our work is 
\begin{equation}
\label{eq:fullconstraint}
    \hat{C} = \hat{F} + \hat{G},
\end{equation}
which is implementable in a computational framework. This enables us solve it using exact numerical methods as well as employing the NNQS ansatz. We note that one now is haunted by the curse of high dimensionality in that one now has a Hilbert space thrice as large as one encounters in the $\Uqone$ model (see Appendix \ref{app:B} for details). In this work, we will also compare the solution space of this master constraint to the Thiemann regularised quantum Hamilton constraint as previously shown in \eqref{eq:trc}. However, the entirety of this discussion, including the computational implementation, is done in Section \ref{sec:results_qhc}.

\subsection{The network architecture}

The main aim of this work is to, of course, solve the presented LQG model using NNQS. However, in this endeavour, we have also taken up the task of developing a network robust enough to solve arbitrary such models. This is due to the fact that it was observed in \cite{Sahlmann:2024pba} that finding a network which can solve such models is a non-trivial task. In this work, we test the capability of the network developed in \cite{Sahlmann:2024pba} and consequently opt to use the same network. 
\newline
The network considered in \cite{Sahlmann:2024pba} was composed of two major blocks, denoted the learning and evaluation blocks respectively. The learning block was a series of convolution sub-blocks which either implemented skip connections or not depending on the dimensions of the Hilbert space. Furthermore, the number of convolution sub-blocks also scaled with the dimensions of the Hilbert space at hand. Each convolution sub-block consisted of several layers starting from a convolutional layer and ending with a pooling layer with several others in between (e.g. normalisation, dropout, activation). The evaluation block was a simple feed-forward network (FFN) the outputs of which were interpreted as the amplitudes for the batches of configurations of basis states given as an input to the network. 
\newline
The scaling of the network was done in effort to maintain a balance between the high accuracy and the computational costs relative to the dimensions of the Hilbert space as it was noticed that for higher $\jmax$, the network needed to be adapted. Currently, this task of maintaining high accuracy has proven to be difficult in this work (for this type of architecture). One encounters several issues contributing to this. First, the graphs are now thrice as large in the computational framework, thus enlarging the size of the input layer of the network. Second, the Hilbert spaces now grow drastically with small increments in $\jmax$. This will introduce a host of issues which will need to be addressed for more complicated models or high $\jmax$ simulations (see Section \ref{sec:betterArchitecture}).
\newline
We note that this aim to develop such a universal network architecture is not needed, as one can develop different networks tailored specifically for different problems. However, as one knows very little about the solution space, in high dimensions as is considered in this work, exact numerical methods become no longer possible due to computational limits. Coupled with the sampler issue discussed in Appendix B of \cite{Sahlmann:2024pba}, one is then inclined to develop a trustworthy network for solving such gravitational models. Establishing trust in a network can be done by verifying its capability across different models. However, this has not been fully achieved here yet. As a result, this means that for high $\jmax$, the accuracy is low compared to what was presented in \cite{Sahlmann:2024pba}. In this work, we only explore low $\jmax$ cutoffs. A brief discussion on possible improvements on the network architecture is presented in Section \ref{sec:betterArchitecture}.

\setcounter{footnote}{0}
\section{Results}
\label{sec:results}
The $\Uqone^3$ BF-theory as described above was implemented in a computational framework where the solutions were obtained using the NNQS ansatz. In this section, we present the results. In Section \ref{sec:results_groundState} we look into the nature of the solutions of the master constraint for different $\jmax$ cutoffs and compare it to the $\Uqone$ model, for validation purposes. Next, we discuss holonomy operators and their fluctuations in Section \ref{sec:results_observables}. In Section \ref{sec:results_volumeOperator} we present a discussion of the spatial volume operator in 3d Euclidean gravity in the $\Uone^3$-limit and its properties. Lastly, in Section \ref{sec:results_qhc}, we present the solutions of the TRC, discuss the nature of the states obtained by the NNQS ansatz and compare them to the solutions of the master constraint $\hat{C}$ \eqref{eq:fullconstraint}.

\subsection{Solutions to the master constraint $\hat{C}$}
\label{sec:results_groundState}
\begin{figure}[h]
    \centering
    \includegraphics[scale=0.3]{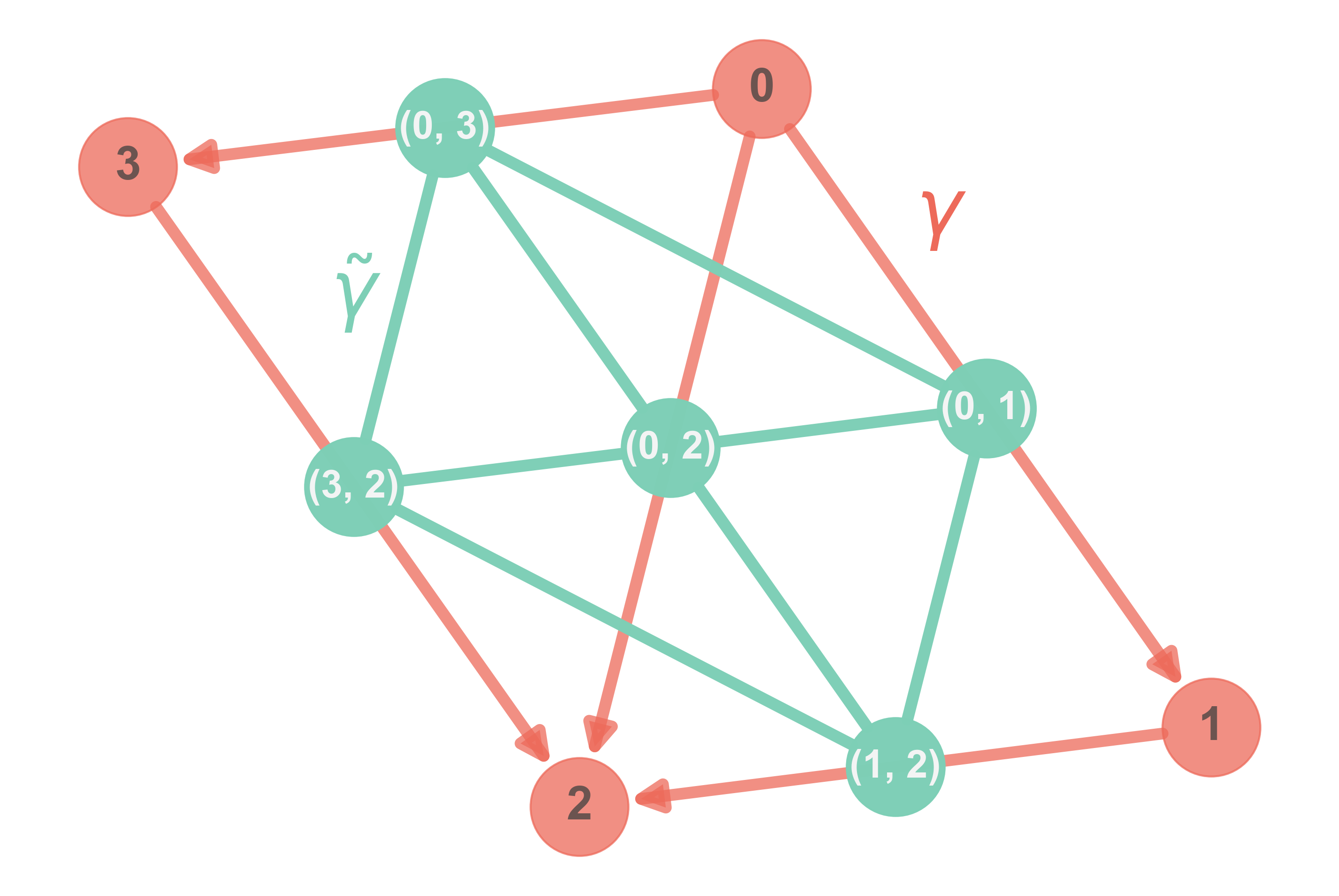}
    \caption{In red, the oriented 2-L graph $\gamma$ is shown and its dual graph $\Tilde{\gamma}$ is shown in green.}
    \label{fig:graph}
\end{figure}

\begin{figure}[h]
    \centering
    \begin{subfigure}{.5\textwidth}
        \centering
        \includegraphics[width=1.0\linewidth]{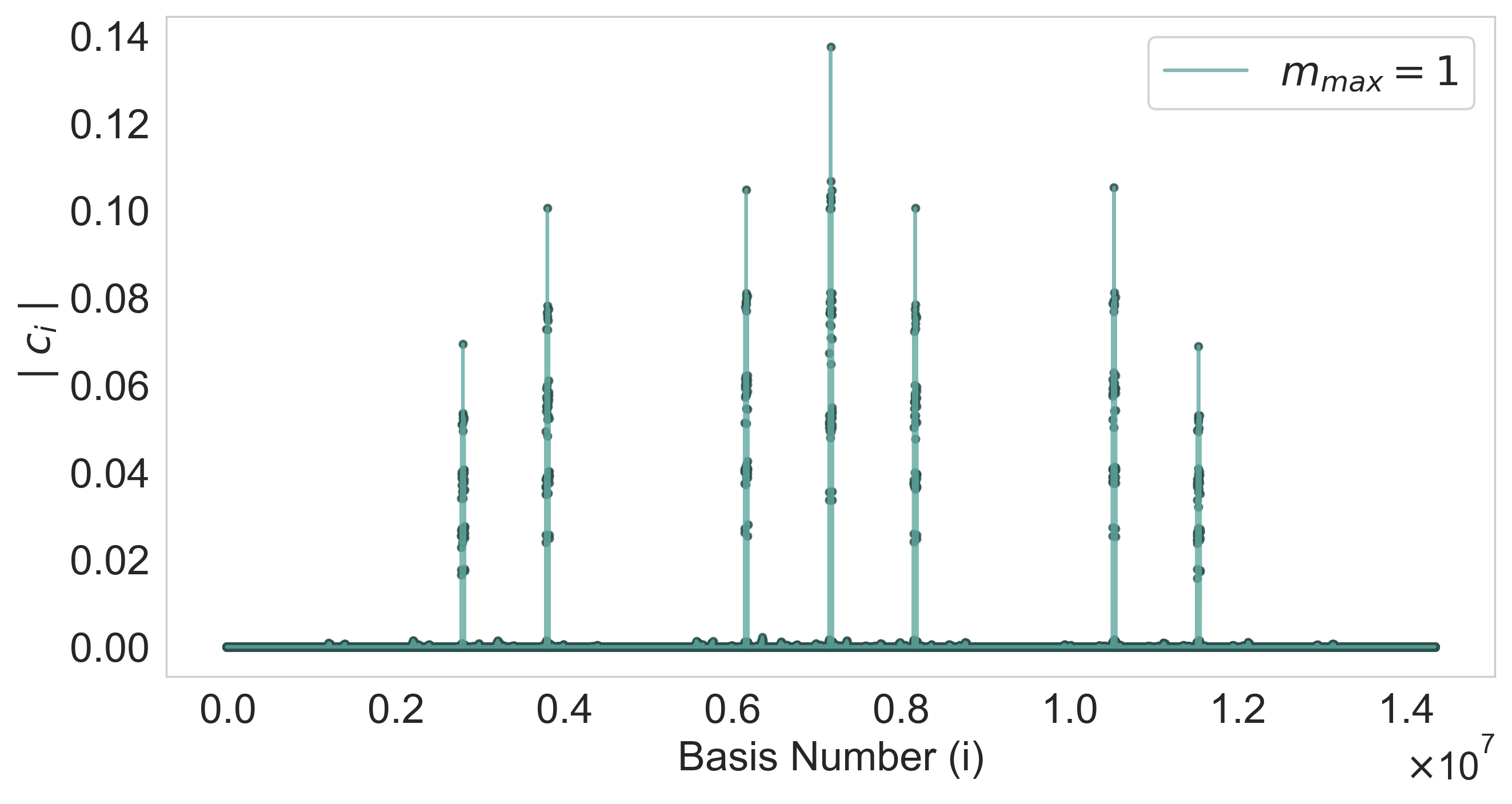}
        \caption{$\Uqone^3$ model}
        \label{fig:s1_groundStateComparison_sub1}
    \end{subfigure}%
    \begin{subfigure}{.5\textwidth}
        \centering
        \includegraphics[width=1.0\linewidth]{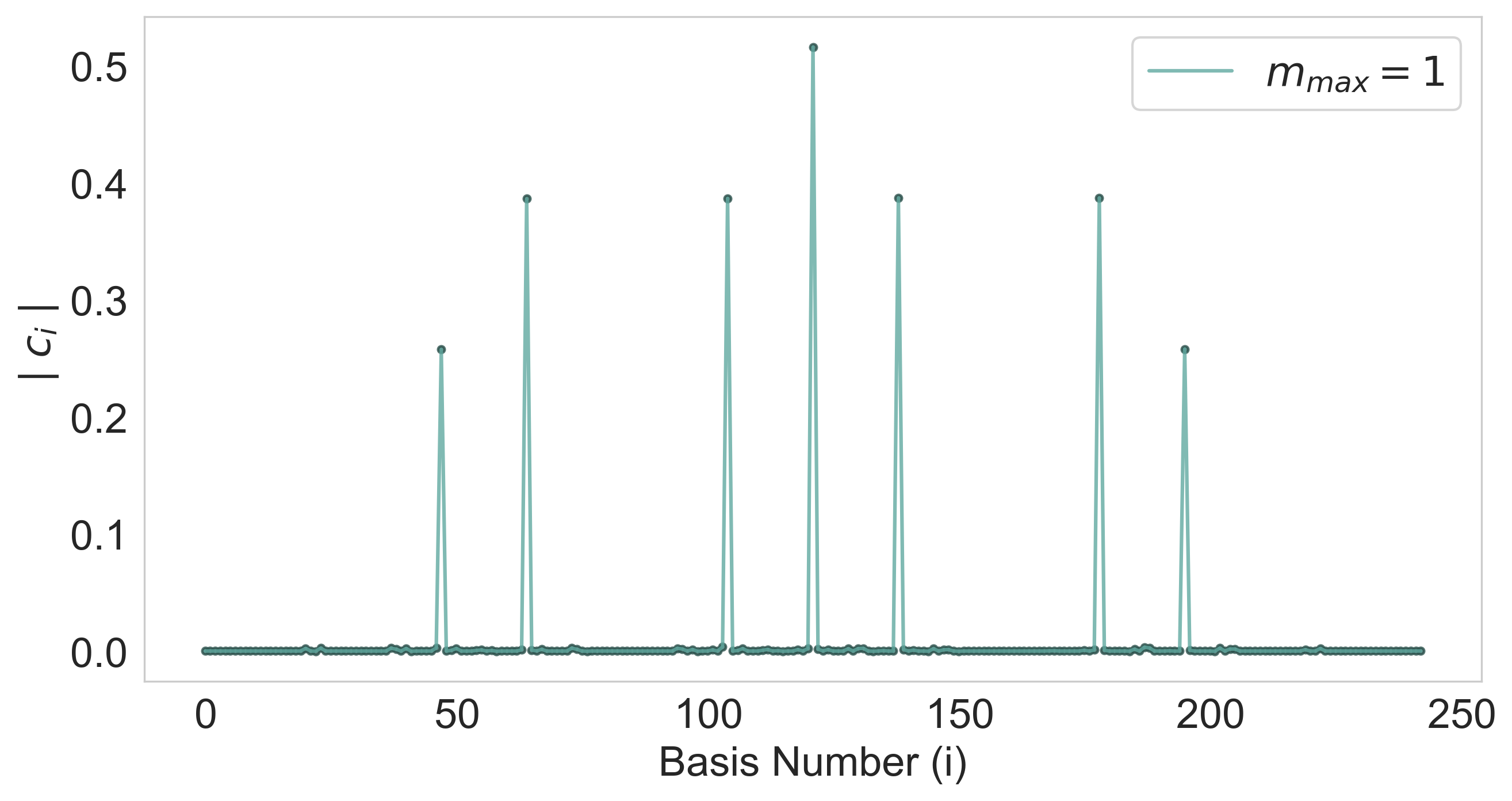}
        \caption{$\Uqone$ model}
        \label{fig:s1_groundStateComparison_sub2}
    \end{subfigure}
    \caption{Solutions to the master constraint of both the $\Uqone^3$ and $\Uqone$ models in the $\jmax = 1$ cutoff are shown. The basis numbering $i \in \mathbb{N}$ is merely a sequential numbering in the computational framework. It is evident that the solution both models at the same cutoff have qualitatively the same features in that the contributing states are similar.}
    \label{fig:s1_groundStateComparison}
\end{figure}

To begin, we define the graph. In previous work \cite{Sahlmann:2024pba}, a specific graph $\gamma$ was considered which was the smallest non-trivial graph with $|L(\gamma)| = 2$. The network used then was able to solve the $\Uqone$ model on $\gamma$ very efficiently. In this work, we will consider the same 2-L graph as done in \cite{Sahlmann:2024pba} which can be seen in Figure \ref{fig:graph}.
\newline
In principle, one can consider any graph. For example, one can consider a graph corresponding to/capturing the topology of a torus, thus solving the model defined on a spacetime with a toroidal spatial geometry. While this may be more physically interesting, it will be left for future work. The reason for considering the graph identical to \cite{Sahlmann:2024pba} is detailed in what follows.
\newline
As mentioned in Section \ref{sec:computationalModel}, the dimensions of the Hilbert space of a $\Uqone^3$ model is thrice as large compared to a $\Uqone$ model defined on the same graph. Thus, this puts serious limitations due to the computational resources required (see Appendix \ref{app:B}). One such limitation is the inability to use exact diagonalisation methods. This, in turn, implies that the results obtained from the NNQS ansatz cannot be numerically verified. However, the structure of both the curvature and Gau{\ss} constraints in the $\Uqone^3$ model correspond to them being the second quantisation operators\footnote{$\hat{A} = \extd\Gamma(\hat{A}^{(1)}) = \hat{A}^{(1)} \otimes \one \otimes \one + \one \otimes \hat{A}^{(1)} \otimes \one + \one \otimes \one \otimes \hat{A}^{(1)}$} of their $\Uqone$ counterparts. As such, the eigenvalues $\lambda$ of the matrix $[\hat{C}]_{ij}$ representing the master constraint $\hat{C}$ in this work is merely $3\lambda^{(1)}$, where $\lambda^{(1)}$ are the eigenvalues of the matrix representing $\Uqone$ constraint. Therefore, to validate the results obtained in this work, one is required to solve the $\Uqone$ model first. This is therefore why we choose to work with the same graph considered in \cite{Sahlmann:2024pba}. \emph{The exact diagonalisation (ED) results shown throughout this work are therefore not computed, but rather taken to be three times the what is obtained in \cite{Sahlmann:2024pba}}. This means that we now test the capability of the NNQS ansatz for a model which one cannot solve using exact methods. This, on its own, is a remarkable point as one is able to explore models which would otherwise be out of reach due to very real computational limits. Lastly, in this work we search for solutions with only real valued coefficients in the charge network basis, which is the computational basis for the numerical work. This is done to avoid the increase in computational demand from complex coefficients \footnote{using complex datatype (\texttt{complex128}) would require double the amount of allocated memory (16 bytes) per variable compared to double precision floating point datatype (\texttt{float64})}. We note that this is a much less drastic restriction than what it seems, since all the constraints have a real valued matrix representation in the computational basis, which ensures that each solution of the constraint can be obtained as a linear combination of solutions with real coefficients 
(see Appendix C in \cite{Sahlmann:2024pba}).
\begin{figure}[h]
    \centering
    \includegraphics[scale=0.45]{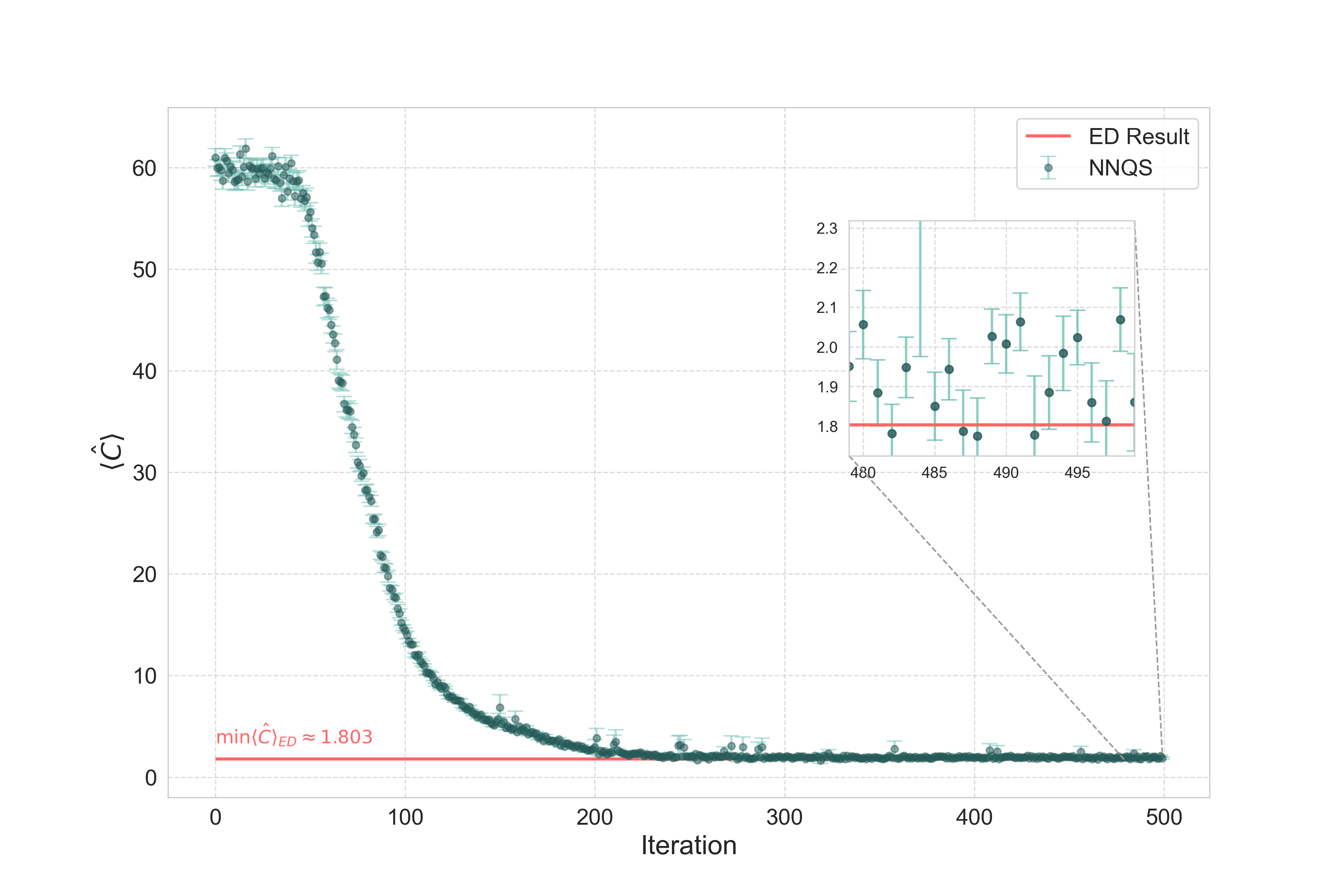}
    \caption{A simulation where the NNQS has been used to solve the $\hat{C}$ constraint in the $\Uqone^3$ model with a cutoff $\jmax = 2$. The accuracy achieved in this simulation was approximately 99.63\%. The number of network parameters needed to be optimised is 6412, constituting only $\sim 2\times 10^{-5}\, \% $ of the dimensions of the Hilbert space ($\sim 3\times 10^{10}$).}
    \label{fig:spin2Run}
\end{figure}
\newline
Furthermore, as shown in Figure \ref{fig:spin2Run}, the network is able to solve such models remarkably well at least for relatively low $\jmax$ cutoffs, as shown in a simulation for a $\jmax = 2$ cutoff where the accuracy achieved was 99.63\%. The accuracy here denotes how close the value of $\min{\expect{\hat{C}}}$ obtained by the network at the final state at the end of the simulation to that of the estimated true value. If we consider the average $\min{\expect{\hat{C}}}$ over the last 100 iterations in the simulation, the accuracy then is approximately 91.38\%. In what follows, by accuracy we denote the former definition rather than the latter. Lastly, the error bars shown in the figure arise due to calculations inherit to the minimisation process, for example the Markov-Chain Monte-Carlo process.
\newline
Also seen in Figure \ref{fig:spin2Run}, the NNQS ansatz with the network architecture used demonstrates the same capabilities as seen in \cite{Sahlmann:2024pba}. Namely, one still obtains the relatively fast convergence due to the convolutional layers and the specific network architecture utilising them. This is made even more remarkable when it is realised that for the case of $\jmax = 4$ as an example, we have only $42101$ network parameters to optimise in order to find the solution. While this may seem large at first, when compared to $\dim\hilb_{\gamma} = 205.89 \times 10^{12}$ it is realised that we only need a vanishingly small amount of information, equivalent to almost $2.045\times 10^{-8}$ \% of the entire space, to obtain the solution and thus performing \textit{extreme} dimensionality reduction and drastically reducing the computational cost. Further, the computation time is rather \quotes{short} even on standard, not high performance computing, hardware\footnote{e.g. for $\jmax = 1$ the $\Uqone$ model requires $\approx$ 7 seconds to solve and a $\Uqone^3$ model requires $\approx$ 30 minutes on a standard commercial 8-core Apple Silicon M1 chip, without multiprocessing}. We do observe more prominent fluctuations and larger error bars in the simulations compared to the simple $\Uqone$ model considered in \cite{Sahlmann:2024pba}. While these are not unusual, they may hint at room for improvement in the used architecture. Further, it was also noticed that for higher $\jmax$ valued, the reliability of the network gradually becomes lower, further indicating the need for a better architecture. The following table summarises the values for $\min\expect{\hat{C}}$ obtained using the NNQS ansatz for different $\jmax$ cutoffs.
\begin{table}[h]

    \centering
    \begin{tabular}{c|ccc}
    \rowcolor{lightergreen!50}
        $\jmax$ & $\min{\expect{\hat{C}}_{(\mathrm{ED})}^{**}}$ & $\min{\expect{\hat{C}}_{(\mathrm{NN)}}}$ & Accuracy (\%) \\
        \hline
        1 & 2.507903 & 2.998 ± 0.017 & 80.441 \\
        2 & 1.803495 & 1.74 ± 0.16 & 96.286 \\
        3 & 1.168658 & 1.12 ± 0.11 & 96.069 \\
        4 & 0.790868 & 0.84 ± 0.21 & 93.788 \\
    \end{tabular}
    \caption{The values of $\min\expect{\hat{C}}$ at different cutoff values are shown. The exact diagonalisation results (ED) are compared to the results from the neural network (NN). Note that values are truncated to 6 decimal values at most. \textsuperscript{**}Here, the ED results are not computed but are estimated based on the results obtained in \cite{Sahlmann:2024pba}.}
    \label{tab:groundstate_results}
\end{table}
\newline
In Table \ref{tab:groundstate_results}, the values of $\min\expect{\hat{C}}$ are obtained using the NNQS ansatz for different $\jmax$ cutoffs. Here, the (ED) results are not computed but estimated based on the values obtained in \cite{Sahlmann:2024pba}. The cutoff ranges considered are relatively small due to the fact that the network begins to perform relatively poorly for higher $\jmax$ (see Section \ref{sec:betterArchitecture} for a discussion on possible improvements).  Nevertheless, for low $\jmax$, one sees that one obtains a good accuracy. Furthermore, one sees once again the effect of imposing the cutoff on the charge vectors in that the curvature and Gau{\ss} constraint terms of $\hat{C}$ no longer commute. Despite that, it is also seen that for higher $\jmax$, this is alleviated and one approaches the continuum theory, just as in the case of the $\Uqone$ model \cite{Sahlmann:2024pba}. 
\newline
Because exact diagonalisation is not possible even for $\jmax = 1$ due to the computational limits mentioned, one cannot calculate the inner product of the ED ground state and its NN counterpart $\langle \Psi_{(\mathrm{NN})} | \Psi_{(\mathrm{ED})}\rangle^2$ as done in \cite{Sahlmann:2024pba}. However, for $\jmax = 1$ simulation, one can visualise the amplitudes of the state obtained using the NNQS ansatz. This is shown in Figure \ref{fig:s1_groundStateComparison}.
\newline
Figure \ref{fig:s1_groundStateComparison} shows the amplitudes of the basis states of the solution for the constraint in both the $\Uqone^3$ model on the left and the $\Uqone$ model on the right. Here, the bases are sequentially numbered by integers in the computational framework\footnote{in a $\Uqone$ model over a dual graph with 5 vertices then the state [-1, -1, -1, -1, -1] in an $\jmax = 1$ cutoff is labeled by 0, [-1, -1, -1, -1, 0] by 1, [-1, -1, -1, -1, 1] by 2, and so on}. It is evident that qualitatively, the basis states which contribute to the solution in the $\Uqone^3$ model correspond to similar basis states, in nature, which contribute to the solution of the constraint in the $\Uqone$ model (while now clearly one has many more\footnote{The number of strongly contributing states is roughly 0.2\% of the space} contributing basis states due to the higher number of degrees of freedom). Thus, while one cannot analyse the contributing basis states as done in \cite{Sahlmann:2024pba}, one can confidently extrapolate, to some degree, from the results obtained therein to the case of the $\Uqone^3$ model considered in this work.

\subsection{Quantum fluctuations of observables}
\label{sec:results_observables}
In the $\Uqone^3$ model, one can, in a similar fashion to the $\Uqone$ model, observe the quantum fluctuations of observables in the theory specifically the minimal loop holonomies $\tr \hat{h}_{\alpha_k}$. Since one has now three copies of the $\Uqone$ model, one expects to see that the expectation values of $\tr \hat{h}_{\alpha_k}$ approach 3 for higher $\jmax$ cutoffs. Furthermore, we define the quantum fluctuations as
\begin{equation}
    \Delta\tr \hat{h}_{\alpha_k} = \langle (\tr \hat{h}_{\alpha_k} + \tr \hat{h}_{\alpha_k}^\dagger)^2 \rangle - \langle \tr \hat{h}_{\alpha_k} + \tr \hat{h}_{\alpha_k}^\dagger \rangle^2.
\end{equation}
In doing so, we can compute these values for every minimal loop in the graph. The results are shown in Table \ref{tab:min_loop_holonomy_results}.
\begin{table}[h]
    \centering
    {\renewcommand{\arraystretch}{1.5}
    \begin{tabular}{c|cccc}
        \rowcolor{lightergreen!50}
        \multirow{2}{*}{\cellcolor{lightergreen!50} Observable} & \multicolumn{4}{c}{\cellcolor{lightergreen!50} Charge Cutoff ($\jmax$)} \\
        
        \cline{2-5} 
        \rowcolor{lightergreen!50}
        \multirow{-2}{*}{\cellcolor{lightergreen!50} Observable}
         & 1 & 2 & 3 & 4 \\
        \hline
        $\expect{\tr \hat{h}_{\alpha_1}}$ & 2.18 & 2.707 & 2.903 & 2.684 \\
        $\Delta \tr \hat{h}_{\alpha_1}$ & 3.928 & 1.664 & 1.04 & 0.184  \\

        \hline
        \multicolumn{5}{c}{} \\[-4.6ex]
        \hline

        $\expect{\tr \hat{h}_{\alpha_2}}$ & 2.299 & 2.752 & 2.945 & 2.661  \\
        $\Delta \tr \hat{h}_{\alpha_2}$ & 3.666 & 1.084 & 0.728 & 1.547 \\
        
        \hline
    \end{tabular}
    }
    \caption{Results of the expectation value and quantum fluctuations of the minimal loop holonomy operators, for every minimal loop, in the solution of $\hat{C}$. The findings align with obtaining flat solutions corresponding to the continuum theory for higher $\jmax$.}
    \label{tab:min_loop_holonomy_results}
\end{table}
\newline
Table \ref{tab:min_loop_holonomy_results} presents the expectation values and the quantum fluctuations of the minimal loop holonomy operators for every minimal loop in the graph. The results presented indicate that for higher $\jmax$ values, then the effect of imposing such a cutoff become less pronounced, leading to the observables having a closer value to 3 as one would expect with quantum fluctuations which slowly die out. This behaviour is in analogy to what was observed in the $\Uqone$ case \cite{Sahlmann:2024pba}. The values of these observables at the $\jmax = 4$ cutoff deviate away from what is expected as a result of decreasing accuracy in the simulation at this cutoff. As such, while for high $\jmax$ the results are more difficult to obtain, one can still confidently extrapolate the results. For relatively low $\jmax$, one starts to see the behavior of the continuum theory, in which we get flat solutions.

\subsection{The volume operator in the $\Uqone^3$-limit}
\label{sec:results_volumeOperator}
The spatial volume operator for 3d Euclidean gravity in the $\Uqone^3$-limit implemented here follows the prescription given in \cite{Thiemann:1997ru}. In principle, one expects to see some known behaviour from such an operator. For example, unlike the volume operator of the 4d theory, the volume operator of 3d gravity does not vanish on 2 and 3-valent vertices so long as there is at least a pair of edges incident at the vertex with linearly independent tangents \cite{Thiemann:1997ru}. The purpose of this section is to affirm such results and demonstrate the ability of implementing and computing such quantum geometric observables in our computational framework. 

\subsubsection{Implementing the quantum volume operator}

\begin{figure}[h]
    \centering
    \includegraphics[scale=0.35]{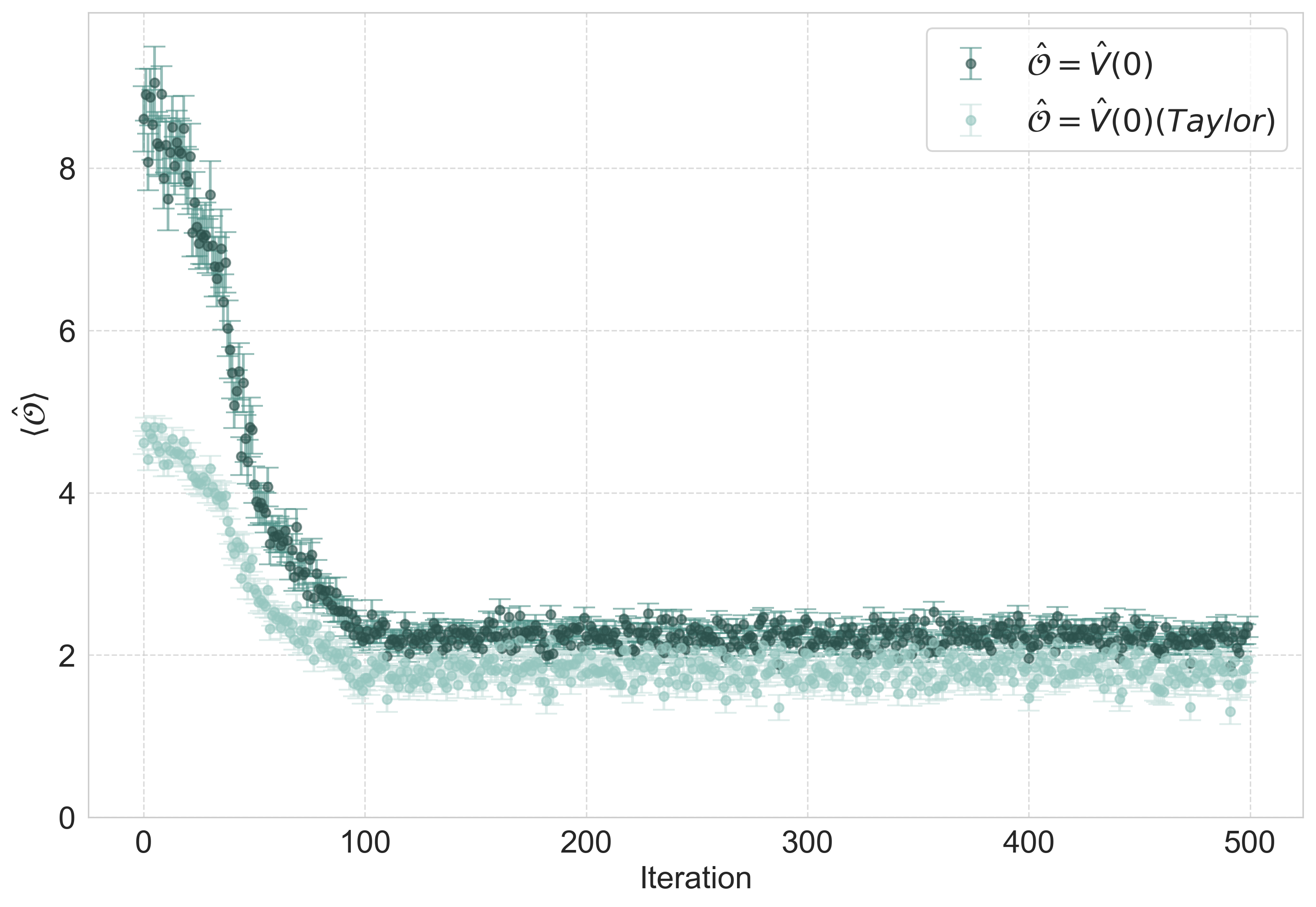}
    \caption{An $\jmax = 1$ simulation is shown and the two implementations of the volume operator are compared. In light green, the Taylor expanded volume operator around $x_0 = 4$ is shown while in dark green, the direct implementation of the volume is shown. As one sees, the Taylor expanded operator gives results that closely match the exact implementation as the simulation reaches a stable solution.}
    \label{fig:comparingVolumeImplementations}
\end{figure}

The discussion starts with the computational implementation of such an operator as it contains subtleties which make it not straightforward. Recall that the vertex contribution  to the volume operator is expressed as \cite{Thiemann:1997ru}
\begin{equation}
\label{eq:volume_}
    \hat{V}_v := \sqrt{\sum_I\left( \sum_{e, e' \text{ at } v} \sign(e, e') \epsilon_{IJK} X^{J}_{e} X^{K}_{e'}\right)^2},
\end{equation}
Here, the $\sign (e, e')$ can be computed directly for every pair of edges at every vertex in the graph. Next, the $X_e$ in our case is the charge vector associated to the edge $e$. Thus, the term $\epsilon^{IJK} X^J_e X^K_{e'}$ is simply the cross product of the two charge vectors of the edges $e$ and $e'$. This cross product is computed element-wise. Once the sum is computed, it can be easily squared. One then faces the issue of implementing the square root, even though the volume operator is a diagonal operator. This is because in our computational framework the operators are stored in a representation which is not in dense form. Rather, there are several \quotes{sub-operators} in the volume operator, each representing a term in the squared sum and acting on a certain number of vertices. The complete volume operator (represented by a $\dim\hilb_{\gamma} \times \dim\hilb_{\gamma}$ matrix) acting on the charge-network function on the entire graph is constructed only at runtime. Even then, only the non-zero matrix elements are stored. This is done to allow for fast computations which do not require large computational resources. Taking the square root of the sub-operators does not correspond to taking the square root of the resulting volume operator. Therefore, the volume operator implemented in our basis is
\begin{equation}
    \hat{V}_{v}^2 := \sum_I\left( \sum_{e, e' \text{ at } v} \sign(e, e') \epsilon_{IJK} X^{J}_{e} X^{K}_{e'}\right)^2.
    \label{eq:volumeSquared}
\end{equation}
From there, we perform a series expansion to obtain an expression for $\hat{V}_{v}$ and once more for $\sqrt{\hat{V}_{v}}$ (see Appendix \ref{app:A}) which is needed later for the TRC. The task of Taylor expanding an operator in the computational framework however introduces further computational complexity, as the operator grows in size for higher order expansions. In order to avoid this, we always expand the volume operator only up to first order. To ensure that such an expansion does not introduce a rather large margin of error, when one intends to use this Taylor expanded volume operator, one is required to run a complete simulation once prior to doing so where the exact operator shown in equation \eqref{eq:volumeSquared} is evaluated at the end for every vertex. This value, for every vertex, becomes the point around which the corresponding operators are Taylor expanded.
\newline
We note, however, that this Taylor expansion is only necessary for using the volume operator in the TRC later. For the case that one wishes to investigate the volume operator alone, we have developed a direct implementation corresponding to the exact expression shown in equation \eqref{eq:volume_}. Consequently, it is important to measure the differences between the two implementations to further determine the margin of error introduced due to the Taylor expansion. The results are shown in Figure \ref{fig:comparingVolumeImplementations}.
\newline
As seen in Figure \ref{fig:comparingVolumeImplementations}, a simulation is conducted in the $\jmax = 1$ cutoff where the $\Uqone^3$ master constraint $\hat{C}$ is being solved and the two different implementations of the  volume operator are being observed. In dark green, the direct implementation of the volume operator $\hat{V}_{v}$ as shown in equation \eqref{eq:volume_} is shown while in light green, the Taylor expanded version of \eqref{eq:volumeSquared} is shown. The Taylor expansion took place around $x_0 = 4$ and both the operators were computed for the same 3-valent vertex in the graph. 
\newline
As can be seen from the figure, the two implementations do not result in the same numerical result. However, the Taylor expansion was not taken to be the most accurate one, as we have deliberately taken a different point to expand around as would be otherwise dictated by the procedure outlined prior. This is to demonstrate that even at first order expansion around a point which is not \textit{very} accurate, the Taylor expanded version of the volume operator does not introduce a large margin of error. More importantly, it shows that the Taylor expanded operator does not exhibit completely different behaviour when acting on the solution of the Master constraint as shown at the end of the simulation in Figure \ref{fig:comparingVolumeImplementations}. This then rules out any room for excluding obtained results due to this approximation. The results in the following section concerning the volume operator are all obtained from the direct implementation rather than the Taylor expanded approximation. This Taylor expansion will only be used in the quantum Hamilton constraint discussed in Section \ref{sec:results_qhc}.

\subsubsection{Properties of the spatial volume operator in 3d Euclidean gravity}

\begin{figure}[h]
    \centering
    \includegraphics[scale=0.35]{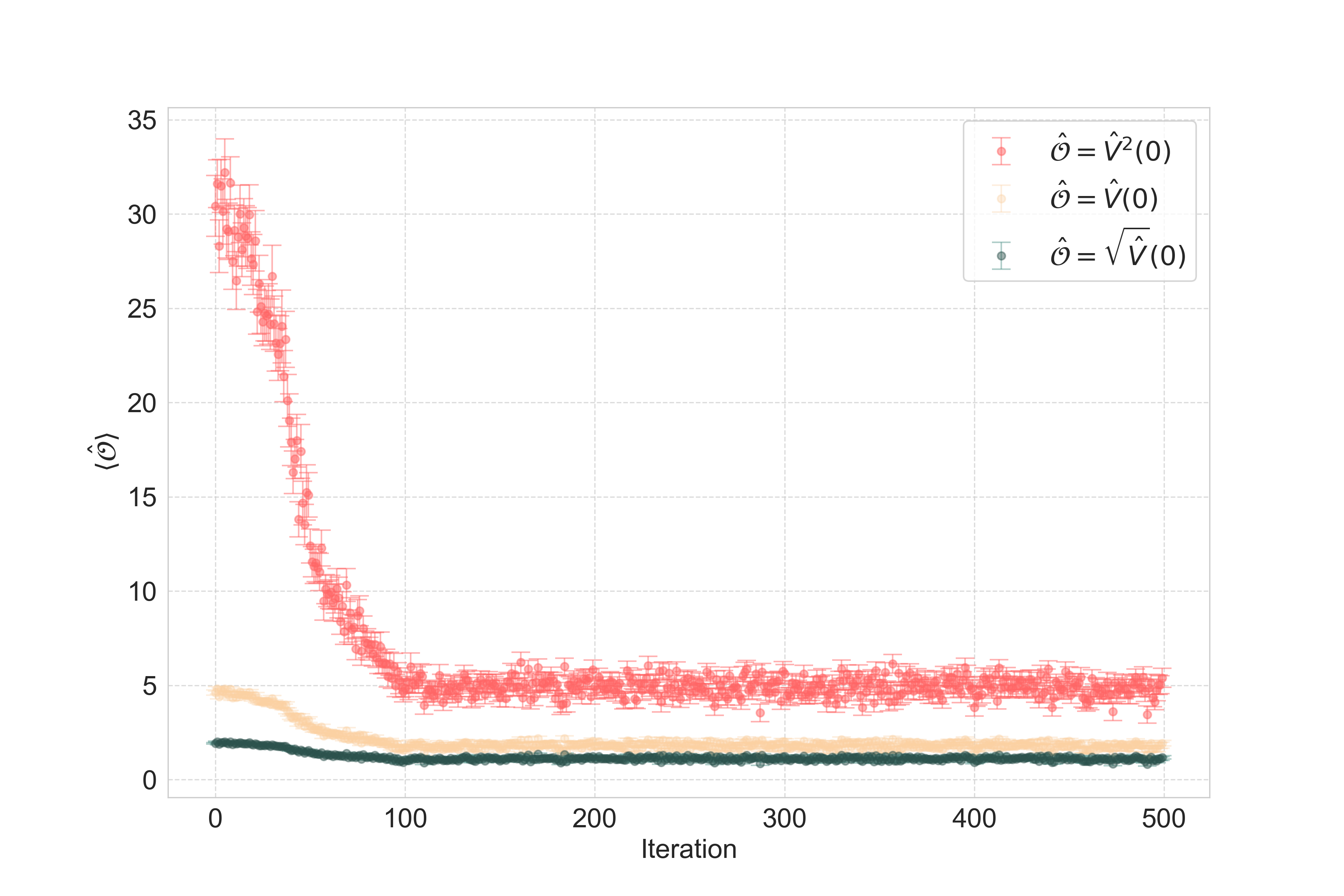}
    \caption{An $\jmax = 1$ simulation where the constraint $\hat{C}$ is being solved and the expectation value of the volume squared, the volume and the square root of the volume operators are observed while acting on a 3-valent vertex of the graph in Figure \ref{fig:graph}.}
    \label{fig:comparingVolumeAndVariants}
\end{figure}

We now shift out attention to discussing the properties of the volume operator in this model. As mentioned, one expects to see that the volume operator not vanishing on 3 or 2-valent vertices identically so long as they have at least a pair of edges with orthogonal tangents at the vertex. Moreover, the Gau{\ss} constraint in our model imposes the $\Uqone^3$ gauge invariance, thus charge conservation at every vertex in each of the graphs. If the Gau{\ss} constraint is well imposed, and as such the solution of the constraint of the model obtained by the NNQS ansatz is $\Uqone^3$ gauge invariant, we expect to see some vertices to have zero volume. Specific to our case, due to the orientation of the graph, we expect to see the 2-valent vertices to have vanishing volume as the charge vectors on the edges attached to them can be completely fixed by the Gau{\ss} constraint leading to their cross product being zero.
\begin{figure}[h]
    \centering
    \includegraphics[scale=0.35]{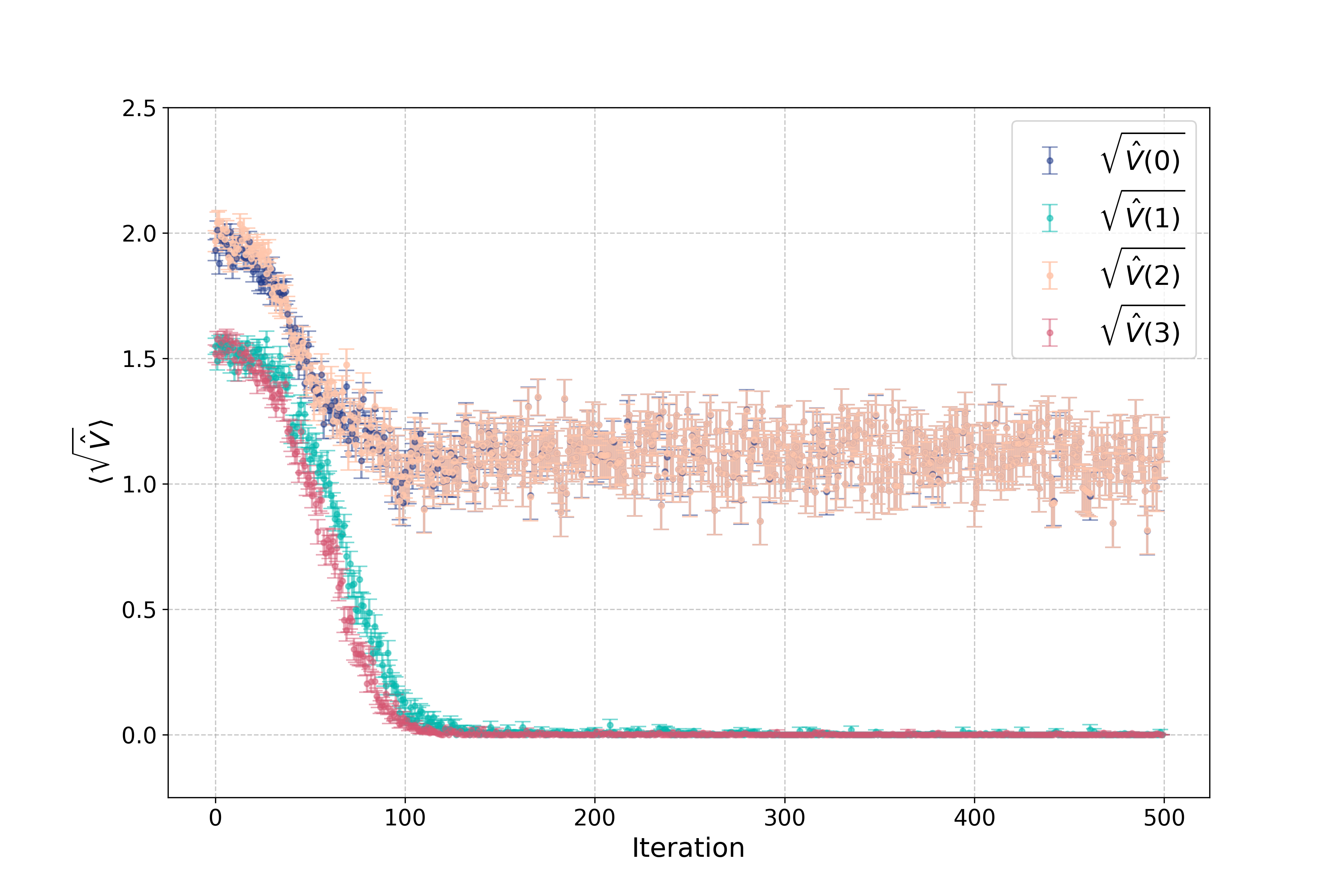}
    \caption{The expectation value of $\sqrt{\hat{V}}$ for every vertex in the graph evaluated during a $\jmax = 1$ simulation where the constraint $\hat{C}$ was solved with an accuracy of 81.013\%. As shown in blue and beige, $\sqrt{\hat{V}}$ does not vanish on 3-valent vertices (labeled by 0 and 2 in Figure \ref{fig:graph}). The 2-valent vertices 1 and 3 have a zero volume as shown since the charge vectors on the edges attached to them are fixed by the Gau{\ss} constraint.}
    \label{fig:volumeResults}
\end{figure}
\newline
Figure \ref{fig:volumeResults} shows the expectation value of $\sqrt{\hat{V}}$ on every vertex of the graph during an $\jmax = 1$ simulation where the constraint $\hat{C}$ was solved with an accuracy of 81.013\%. As can be seen, the 3-valent vertices labeled in our graph by 0 and 2 do not have vanishing volume while the 2-valent vertices do. This is because the solution of the constraint we arrive at has contributing basis states all of which are gauge invariant. Thus the Gau{\ss} constraint, as expected, perfectly fixes the charge vectors on the edges attached to these 2-valent vertices and as a result, the volume operator, which involves the cross product of now two identical vectors, becomes zero. This holds true even for higher $\jmax$, with the only difference being the amount of numerical instability and fluctuations in the simulation becoming lower. This is the result of more allowed charge configurations being allowed.
\newline
We now look at one last result concerning the volume operators. Namely, what can be learned about the nature of the solution from the volume operator. Figure \ref{fig:comparingVolumeAndVariants} shows the expectation value of the volume squared, the volume, and the square root of the volume operator all evaluated on the 3-valent vertex labeled by 0 in our graph during a $\jmax = 1$ simulation where once again the constraint $\hat{C}$ was solved. 
\newline
As shown in Figure \ref{fig:comparingVolumeAndVariants}, the $\expect{\sqrt{\hat{V}}}$ on the vertex 0 is close to the square root of $\expect{\hat{V}}$, which in turn is close to the square root of $\expect{\hat{V}^2}$. Specifically, at the end of the simulation, the values were $\expect{\hat{V}^2} = 4.69 \pm 0.43$, $\expect{\hat{V}} = 1.78 \pm 0.15$ and $\expect{\sqrt{\hat{V}}} = 1.112 \pm 0.088$. One can see that $\expect{\hat{V}} \approx \sqrt{\expect{\hat{V}^2}}$ but not \textit{exactly} equal and similarly, $\expect{\sqrt{\hat{V}}} \approx \sqrt{\expect{\hat{V}}}$\footnote{$\sqrt{4.69} \approx 2.165, \sqrt{\sqrt{4.69}} \approx 1.471$}. Now, consider the expression for the expectation value of a function $f(\hat{V})$ of the volume operator denoted as follows
\begin{equation}
    \expect{f(\hat{V})} = \int_{\mathrm{Spec}\hat{V}} p(\lambda) f(\lambda) d\lambda.
\end{equation}
Let us now interpret $p(\lambda)$ as the probability distribution of the wave-function. If one finds that this distribution is sharply peaked around a specific value $\lambda = v_0$, then the expectation value would simply be $\expect{f(\hat{V})} = f(v_0)$. In the other extreme, if it is evenly spread this equality would not hold. In the case that the distribution is sharply peaked, one has another equality which holds true, namely $\expect{\sqrt{f(\hat{V})}} = \sqrt{\expect{f(\hat{V})}}$. The findings presented above suggest that the ground state obtained using the NNQS ansatz is not as sharply peaked as a $\delta$-distribution in the spectrum of the volume operator.

\setcounter{footnote}{0}

\subsection{Investigating the Thiemann regularised quantum Hamilton constraint of 3d LQG}
\label{sec:results_qhc}
So far, we have explored the solution space of the master constraint $\hat{C}$ composed of a curvature part $\hat{F}$ and a Gau{\ss} part $\hat{G}$. In this section, we investigate the solution of the quantum Hamilton constraint constructed using the Thiemann strategy \cite{Thiemann:1997ru}. We first discuss the implementation of such a quantum Hamilton constraint on the computational framework. Following that, we present several results starting with observing the behaviour of the TRC in the solution space of $\hat{C}$. We follow that with exploring states near the kernel of the TRC using the NNQS ansatz. Lastly, we compare these states with the solution space of $\hat{C}$.

\subsubsection{Implementing the TRC: challenges and possibilities}
Recall that the TRC of 3d LQG, following the regularisation procedure of \cite{Thiemann:1997ru}, can be expressed as 
\begin{align}
    \hat{H}_T(N) f_\gamma & := \lim_{\epsilon \rightarrow 0} \hat{H}_{T, \epsilon}(N) f_\gamma \\
    & = \frac{2}{\hbar^2}\sum_{\Delta, \Delta' \in T, v} \epsilon^{ij}\epsilon^{kl}N(v) \tr(\hat{h}_{\alpha_{ij}(\Delta')} \hat{h}_{s_{k}(\Delta)} [\hat{h}_{s_{k}(\Delta)}^{-1}, \sqrt{\hat{V}_v}] \hat{h}_{s_{l}(\Delta)} [\hat{h}_{s_{l}(\Delta)}^{-1}, \sqrt{\hat{V}_v}]) f_\gamma.
    \label{eq:trc_}
\end{align}
As described above, we modify he prescription from \cite{Thiemann:1997ru} by letting 
$\alpha_{ij}$ be a minimal loop in $\gamma$, with basepoint $v$ that starts with edge $e_i$ and ends with edge $e_j$. In the graph considered, this means that the triangulation $T$ is simply the set of minimal loops $L(\gamma)$ of the graph\footnote{For a graph composed of minimal loops of length $\geq 3$, then one can adopt the same triangulation by employing edge refinement.}. Therefore, the sum goes over all ordered pairs of minimal loops in the graph, for which then $\epsilon^{ij}$ and $\epsilon^{kl}$ are computed accordingly. We will  drop the numerical prefactor in equation \eqref{eq:trc_} and set $N(v) = 1$. Lastly, the trace in this case translates to having the constraint decompose over the three different graphs in the computational framework. As such, one has for each term in the sum shown in \eqref{eq:trc_} three terms where each one lives on a different graph.
\newline
As one now in the computational framework has three of such an expression \eqref{eq:trc_}, the sizes of the matrices of the sub-operators packed in the TRC are rather large, and grow quickly for higher charge cutoffs. Even if one simplifies the expression analytically, one still has an abundant number of computational operators which need to be tensor multiplied in order to act appropriately on the wave-function defined on $\hilb_{\gamma}$. This issue is further compounded as one now uses the Taylor expanded volume operator. Simply put, one has several computational sub-operators in the TRC stored in the computational framework which themselves are large matrices which all need to be multiplied together. This introduces a computational demand that is rather large for constructing such an operator. For example, even at a cutoff of $\jmax = 2$, then one would need \textit{hours} to merely construct the TRC and a borderline comical amount of RAM (approx. 7.451 Zetta Bytes ($7.451 \times 10^{9}$ Tera Bytes) for a sparse representation of the constraint matrix) in the process of exactly solving it\footnote{One can in principle conduct out-of-core distributed computations but this would nevertheless require large amount of resources.}.
\newline
Furthermore, if one desires to solve the constraint using the NNQS ansatz, one needs to unpack such an operator in a specific manner in the computational framework where one ultimately searches for all non-zero matrix elements of the operator. This process is also computationally expensive when the operator is non-diagonal. The TRC was constructed over a 2-state model ($M = \left\lbrace -1, 1\right\rbrace$) over the same 2-L graph considered in this work. In such a case, one can allocate enough computational resources to inspect the operator in its entirety. It was found that the matrix representing the constraint was not only not diagonal, but \textit{extremely} ill-conditioned to the point of being almost singular. While the case can be, and probably is, different for a model with more allowed states, it nevertheless shows that the constraint is very difficult to implement in the computational framework and difficult to explore in high $\jmax$ cutoffs.
\newline
In any model considered, irrespective of the number of allowed states due to the charge cutoff, it was observed that the TRC is non-Hermitian. This poses yet another computational hurdle as the tools used in this work to employ the NNQS ansatz do not (yet) work reliably or easily for non-Hermitian operators. All together, this now leaves us in a corner: in order to know \textit{anything} about this constraint, we can only currently do simulations in the $\jmax = 1$ cutoff. In this work, we will consider the Hermitian constraint 
\begin{equation}
\hat{C}_{\mathrm{TRC}} = \hat{H} + \hat{H}^\dagger + \hat{G}.
\end{equation}
While one can also consider the Hermitian and positive operator $\hat{H}\hat{H}^\dagger + \hat{G}$, we opt not to do so as the process of computing $\hat{H}\hat{H}^\dagger$ would result in even more computational demands. We have explicitly checked that the operator $\hat{H} + \hat{H}^\dagger$ is not positive in the 2-state model. In that case,  $\hat{H} + \hat{H}^\dagger$ had an eigenvalue spectrum that was symmetric around 0 and a degenerate kernel. After introducing the Gau{\ss} constraint, the eigenvalue spectrum does not have the mirror symmetry as before, and the degeneracy in the spectrum was lifted. Therefore we do not have reason to believe that the operator is positive for models with higher $\jmax$. 
\newline
It is interesting to note that in 4d gravity, the TRC actually takes a simpler form \cite{Thiemann:1996aw}. Although computational hurdles will always exist, the specific hurdles we encounter in the 3d case will not going to be as pronounced in 4d.

\subsubsection{The behaviour of $\hat{C}_{\mathrm{TRC}}$ in the solution space of $\hat{C}$}

The first point of exploration of $\hat{C}_{\mathrm{TRC}}$ is to observe the constraint in the solution space of the master constraint $\hat{C}$. As shown, the current network architecture has proved capable of solving the latter with good accuracy for different charge cutoff values. However, due to the technical difficulties of implementing $\hat{C}_{\mathrm{TRC}}$ as previously outlined, we are restricted in working in the $\jmax = 1$ cutoff. 
\begin{figure}[h]
    \centering
    \includegraphics[scale=0.35]{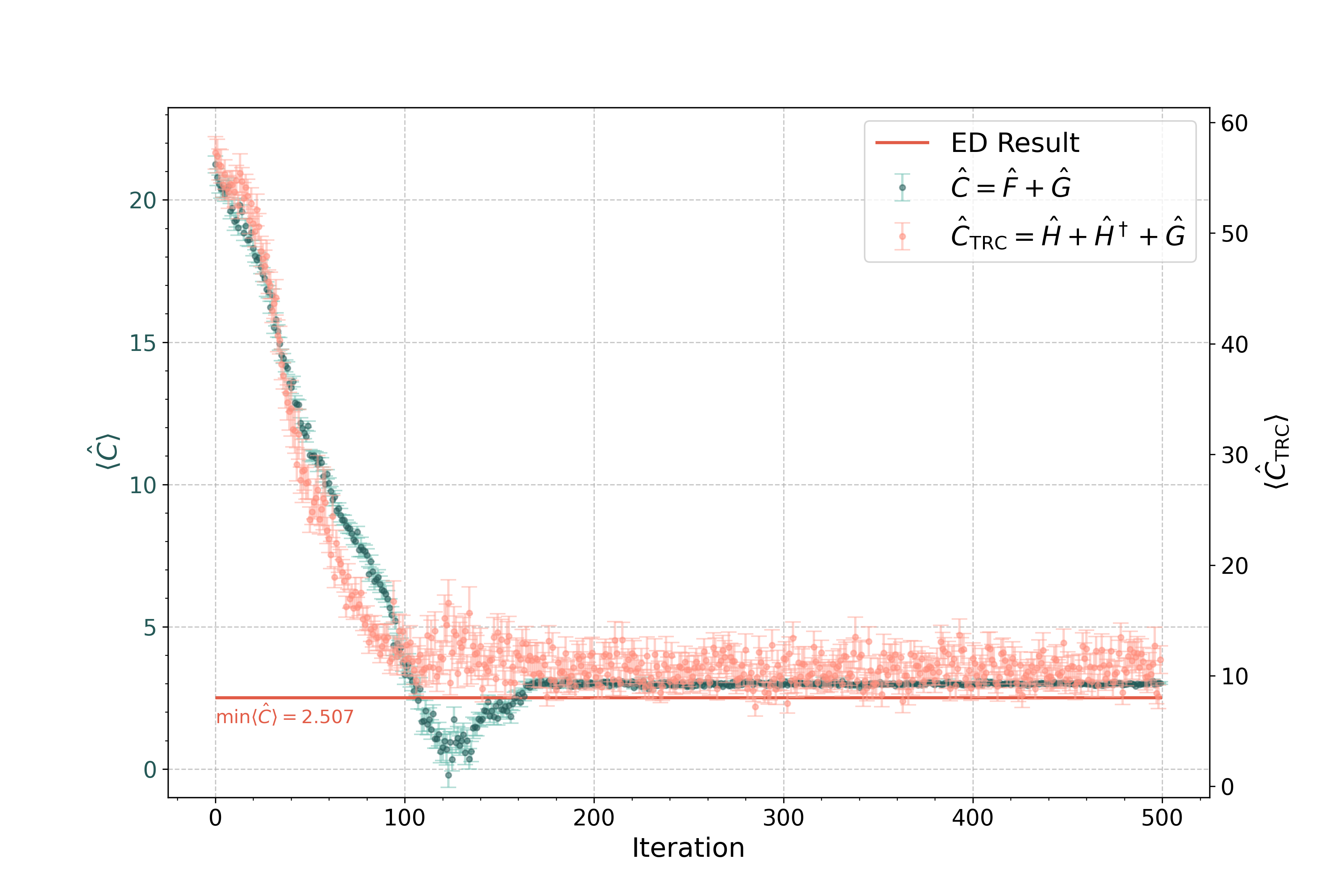}
    \caption{The result of an $\jmax = 1$ simulation where the master constraint $\hat{C}$ is being solved and the constraint $\hat{C}_{\mathrm{TRC}} = \hat{H} + \hat{H}^\dagger + \hat{G}$ is being observed.}
    \label{fig:FGSim_TRC}
\end{figure}
\newline
Figure \ref{fig:FGSim_TRC} shows an $\jmax = 1$ simulation where the master constraint $\hat{C}$, shown in green, is being solved and the TRC constraint $\hat{C}_{\mathrm{TRC}}$, shown in orange, is being observed at every iteration. As one does not know what the solution space of $\hat{C}_{\mathrm{TRC}}$ looks like, it is difficult to conclude whether it is the case that the states that solve $\hat{C}$ also solve $\hat{C}_{\mathrm{TRC}}$. However, it is seen that the solutions of $\hat{C}$ are not in the kernel of $\hat{C}_{\mathrm{TRC}}$. Since exact numerical methods cannot be employed, the only means available to investigate the solution space of $\hat{C}_{\mathrm{TRC}}$ is using the NNQS ansatz. Note that values of $\expect{\hat{C}}$ in Figure \ref{fig:FGSim_TRC} above which are less than 0 are due to numerical artifacts, as further supported by the error bars being large as well, as this mathematically is known to be not possible due to $\hat{C}$ being a positive operator.

\subsubsection{States near the kernel of $\hat{C}_{\mathrm{TRC}}$}

In this section, we attempt to solve the $\hat{C}_{\mathrm{TRC}}$ constraint using the NNQS ansatz. We use the word \quotes{attempt} for the following reasons: the network architecture used here was first developed for $\Uqone$ models, and further shown to be able to solve also $\Uqone^3$ models, despite exhibiting some behaviour indicative of it being somewhat pushed to its limits. The master constraints considered for these theories had a specific expression and was composed of curvature and Gau{\ss} parts. Thus, while the network works well for $\Uqone$ and $\Uqone^3$ models and the Gau{\ss} term is identical in both $\hat{C}_{\mathrm{TRC}}$ and $\hat{C}$, \textit{there is no reason for the network to work well when solving the $\hat{C}_{\mathrm{TRC}}$ constraint as $\hat{H}$ has a completely different mathematical structure compared to $\hat{F}$}. 
\newline
Additionally, the $\hat{C}_{\mathrm{TRC}}$ constraint poses many technical hurdles which severely limits our possibilities to explore it numerically. Thus, aside from the two state model, we do not know what to expect from the eigenvalue spectrum of $\hat{C}_{\mathrm{TRC}}$. Moreover, this operator does not exist in the $\Uqone$ theory, and hence one is left completely in the dark regarding what is to be expected. For this reason, despite the results in this section and the ones which follow being based on several simulations which were conducted using the same parameters to ensure the validity and consistency of the results, there were occurrences where the NNQS has arrived at other solutions. While these occurrences were not abundant, they nevertheless occurred. The results shown here are of one state near the kernel of $\hat{C}_{\mathrm{TRC}}$. However, other states also near the kernel were found to exhibit similar behaviour (see end of Section \ref{sec:comparingSolutionSpaces} for a brief discussion).
\begin{figure}[h]
    \centering
    \centering
    \begin{subfigure}{.5\textwidth}
        \centering
        \includegraphics[width=1.1\linewidth]{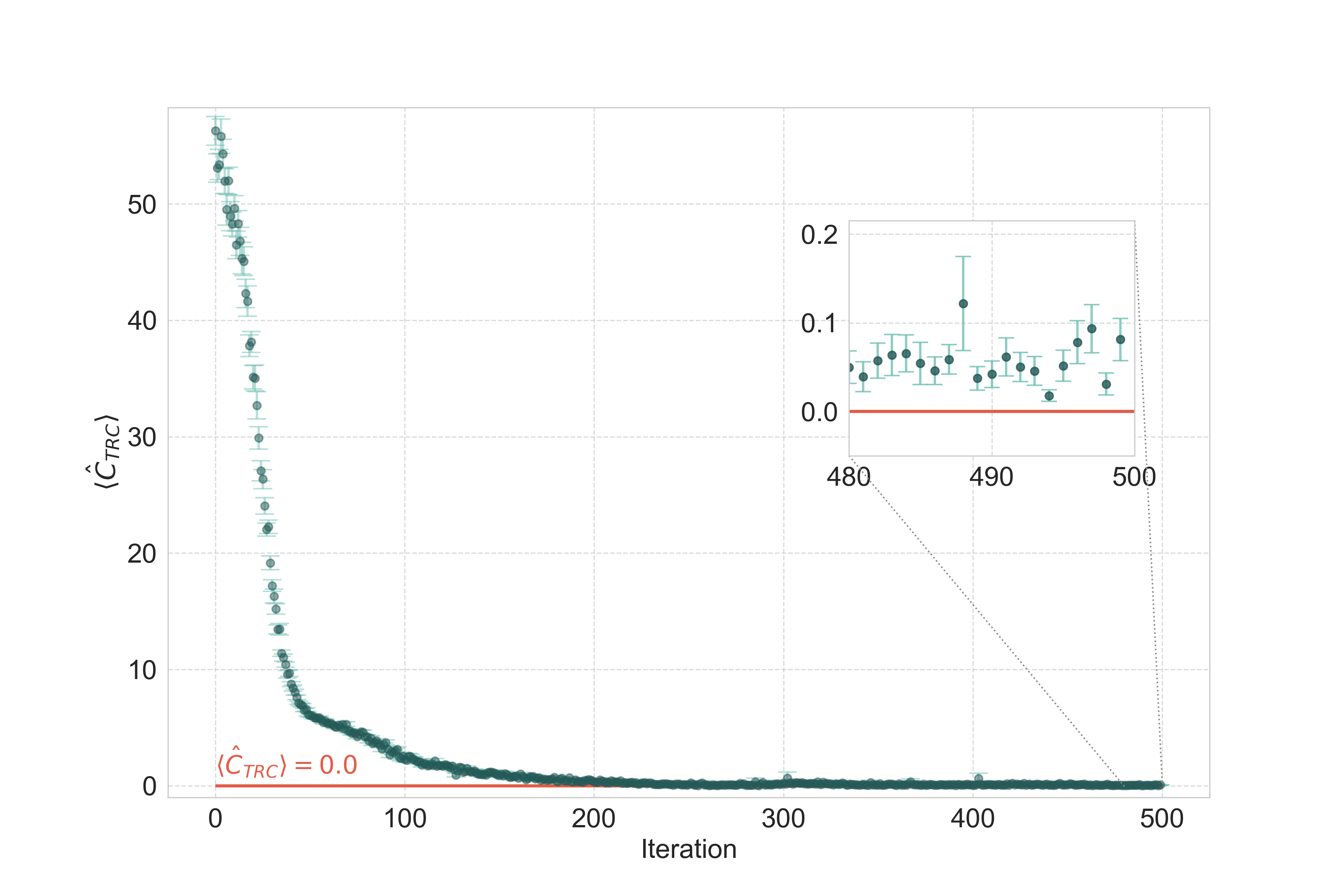}
        \caption{}
        \label{fig:trcResults_sub1}
    \end{subfigure}%
    \begin{subfigure}{.5\textwidth}
        \centering
        \includegraphics[width=1.1\linewidth]{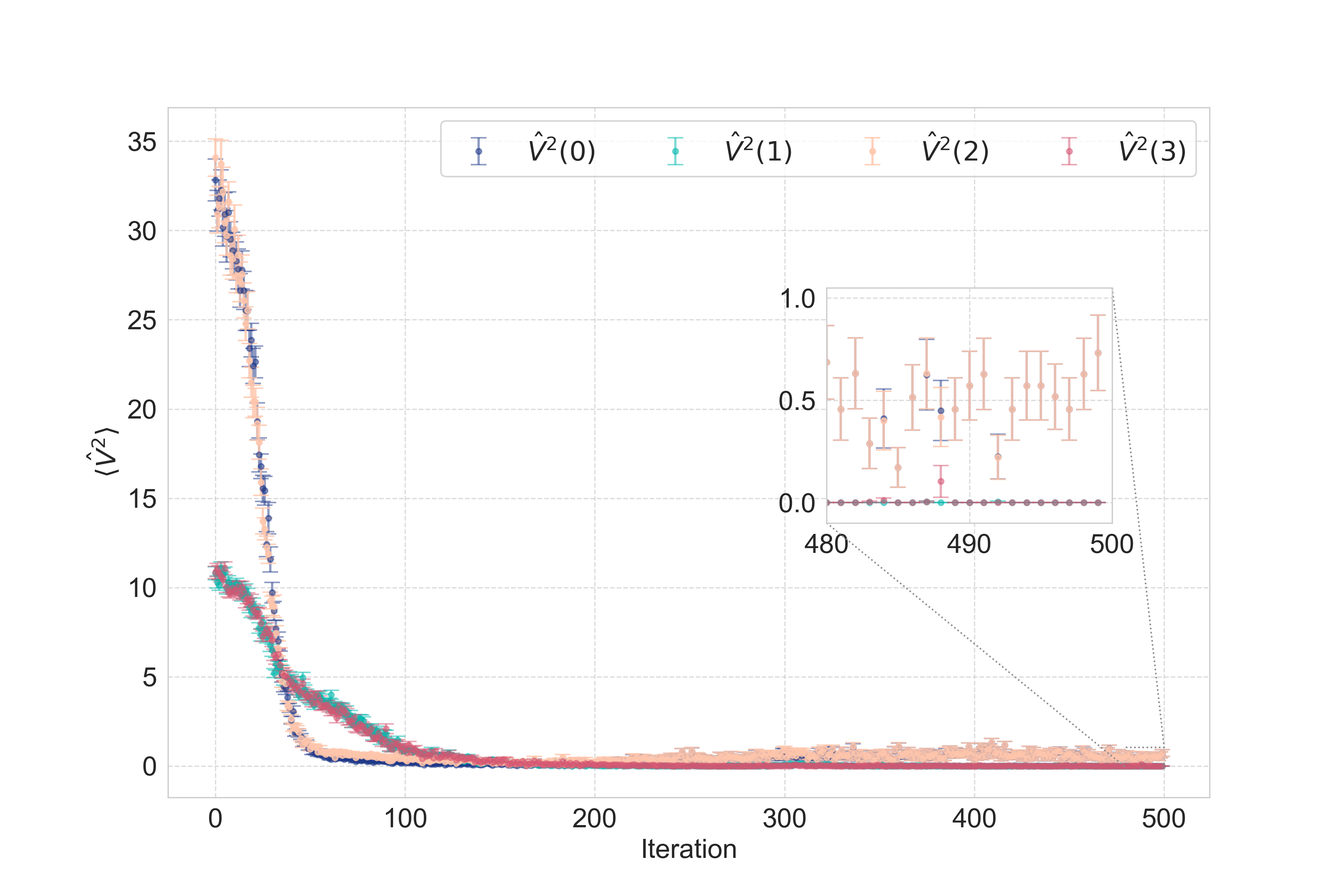}
        \caption{}
        \label{fig:trcResults_sub2}
    \end{subfigure}
    \caption{An $\jmax = 1$ simulation where the constraint $\hat{C}_{\mathrm{TRC}} = \hat{H} + \hat{H}^\dagger + \hat{G}$ is being solved. On the left, we see that by the end of the simulation, we have reached a state almost at the kernel of $\hat{C}_{\mathrm{TRC}}$. On the right, we observe the behaviour of the squared volume operator during the simulation acting on all the vertices of the graph.}
    \label{fig:trcResults}
\end{figure}
\newline
Figure \ref{fig:trcResults} shows a $\jmax = 1$ simulation where the constraint being solved is $\hat{C}_{\mathrm{TRC}}$ using the same network architecture used throughout this work. On the left, the results of the search for the solution space is shown while on the right the volume operator squared is observed on every vertex of the graph during the same simulation. Since the $\hat{C}_{\mathrm{TRC}}$ constraint in the two state model had an eigenvalue spectrum which included negative values, there is no reason a priori to conclude that the three state model is a positive operator. Hence, one is tempted to believe that the state arrived at in the Figure \ref{fig:trcResults_sub1} is not the actual \quotes{ground state} of $\hat{C}_{\mathrm{TRC}}$.
\newline
Nevertheless, as seen in Figure \ref{fig:trcResults_sub1}, we arrive at a state which lies near the kernel of the constraint. Specifically, at the end of this simulation, the state arrived at was one in the eigenspace of value $0.023 \pm 0.012$, close to the kernel. This, actually, is the sort of state one is interested in rather than the actual ground state which may have a negative eigenvalue. Thus, the network used to solve the $\Uqone^3$ model, when employed to solve the $\hat{C}_{\mathrm{TRC}}$ constraint in the $\jmax = 1$ cutoff, remarkably lands us near the kernel of $\hat{C}_{\mathrm{TRC}}$. 
\newline
During the simulation, other operators were also observed, for example the squared volume operator as shown in equation \eqref{eq:volumeSquared}. The results are shown on the right in Figure \ref{fig:trcResults_sub2}. During the simulation, for this specific state, the volume squared operators were observed to behave in a similar qualitative manner as to what was observed in Figure \ref{fig:volumeResults} in the case of solving the $\hat{C}$ constraint. That is, it only vanishes on the same 2-valent vertices. The numerical values are now however much smaller. At the end of the simulation, the squared volumes were observed to have the values $0.732 \pm 0.183$ for both of the two 3-valent vertices here labeled by the numbers 0 and 2 while being zero for the other two 2-valent vertices.
\newline
The last operator which was observed during this simulation is the master constraint $\hat{C}$. This now asks a similar question to that of the previous section, which is whether or not the kernel of $\hat{C}_{\mathrm{TRC}}$ corresponds to the solution space of $\hat{C}$. 
\begin{figure}[h]
    \centering
    \includegraphics[scale=0.35]{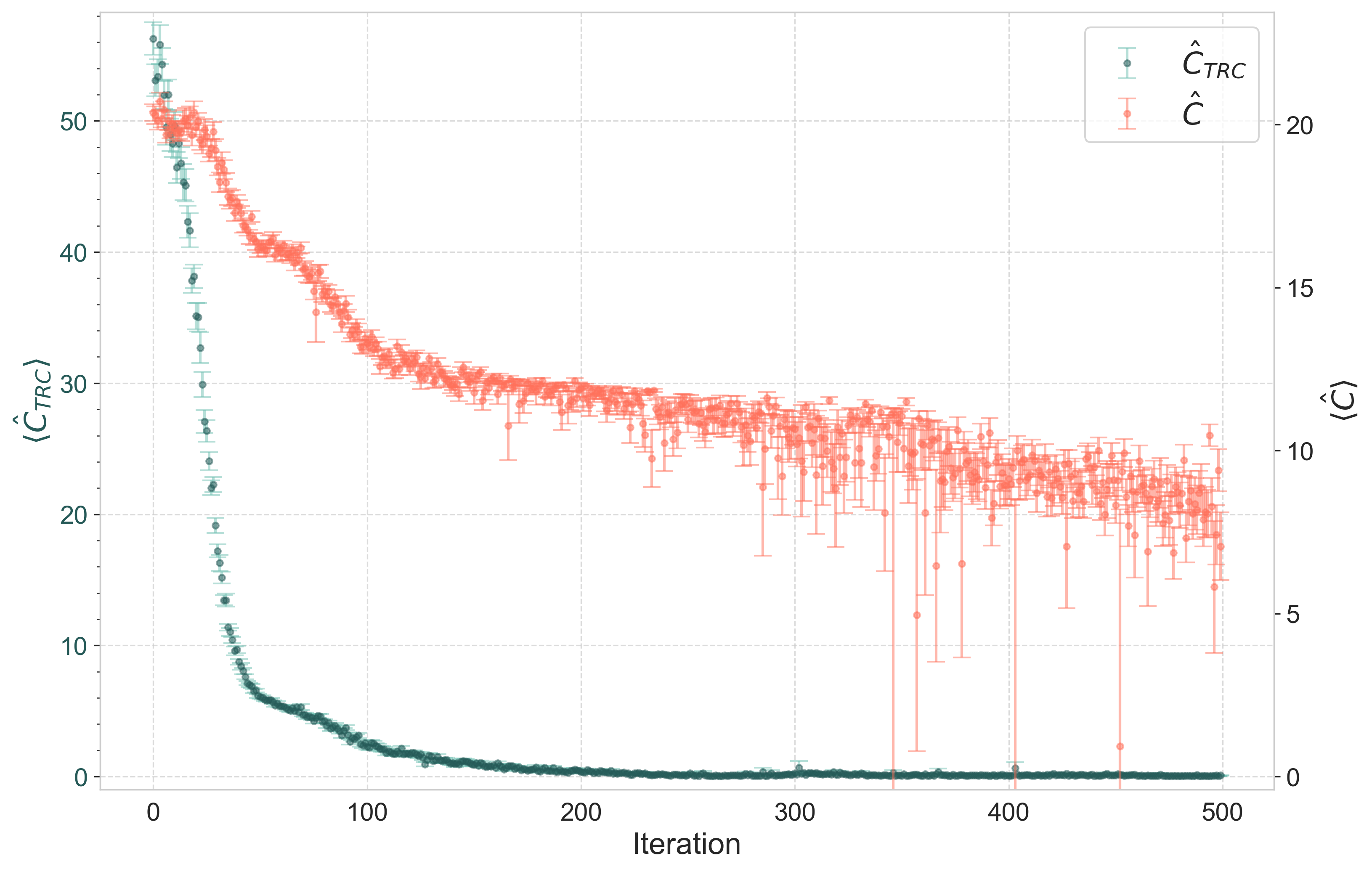}
    \caption{A $\jmax = 1$ simulation where $\hat{C}_{\mathrm{TRC}}$ is being solved, shown in green, and $\hat{C}$ is being observed, shown in orange. As shown, the states which are (almost) in the kernel of $\hat{C}_{\mathrm{TRC}}$ do not correspond to the ground state of $\hat{C}$.}
    \label{fig:HGSim_FG}
\end{figure}
\newline
In Figure \ref{fig:HGSim_FG}, a simulation in the $\jmax = 1$ cutoff is conducted where the constraint being solved is the $\hat{C}_{\mathrm{TRC}}$ constraint as shown in green. The states at the end of the simulation lie almost in the kernel of $\hat{C}_{\mathrm{TRC}}$ as previously discussed. However, one observes that these states do not put us in the solution space of $\hat{C}$.
This is known because (a) in the previous section it was shown that the solution space of $\hat{C}$ does not correspond to the kernel of $\hat{C}_{\mathrm{TRC}}$ and (b), it is further affirmed now, as it is known from Section \ref{sec:results_groundState} that the solution space of $\hat{C}$ in the $\jmax = 1$ cutoff is not the same eigenspace arrived at in Figure \ref{fig:HGSim_FG} above. We now aim to quantify the degree of similarity of these solution spaces.

\subsubsection{Comparing solution spaces of $\hat{C}$ and $\hat{C}_{\mathrm{TRC}}$}
\label{sec:comparingSolutionSpaces}
\begin{figure}[h]
    \centering
    \begin{subfigure}{.5\textwidth}
        \centering
        \includegraphics[width=1.0\linewidth]{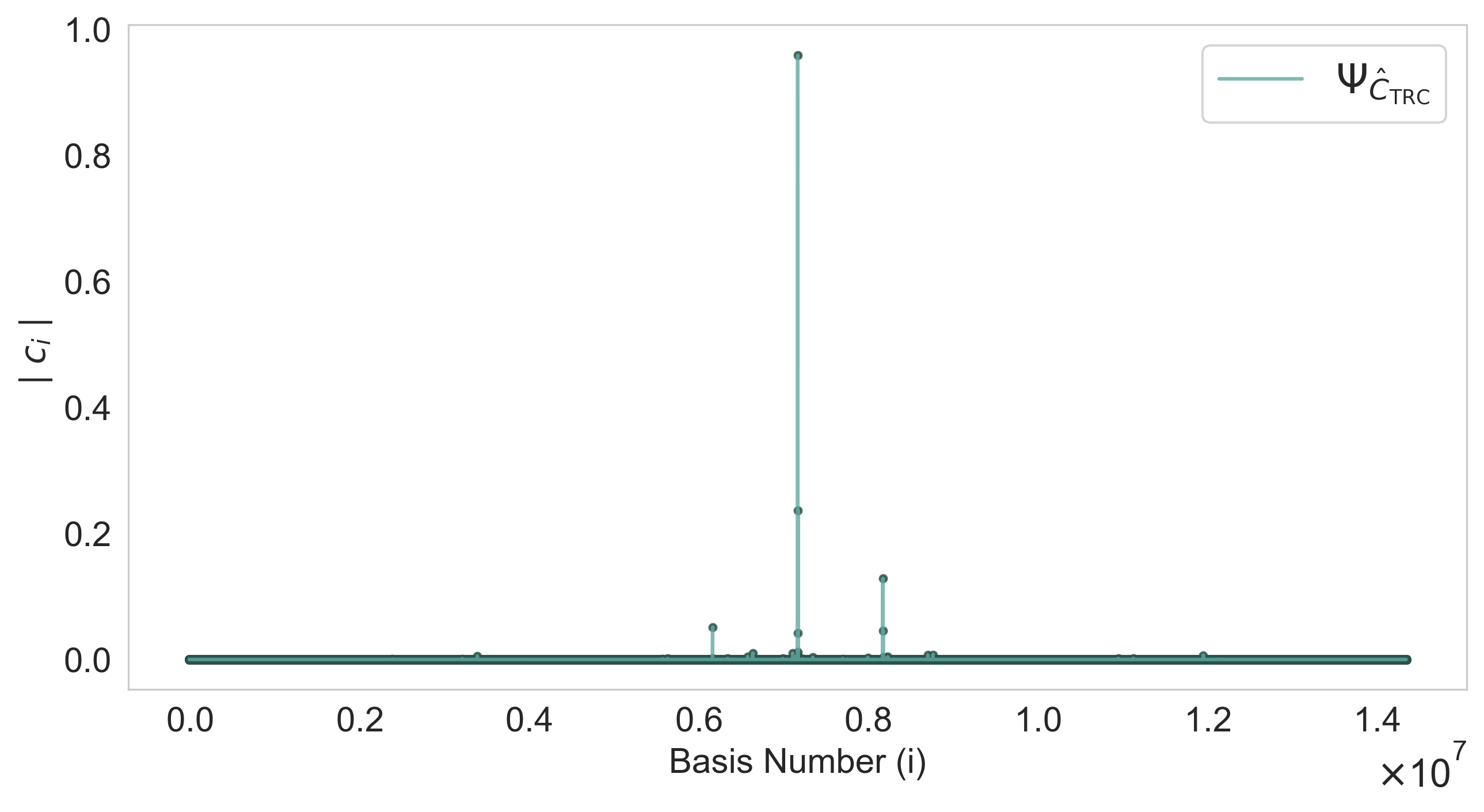}
    \end{subfigure}%
    \begin{subfigure}{.5\textwidth}
        \centering
        \includegraphics[width=1.0\linewidth]{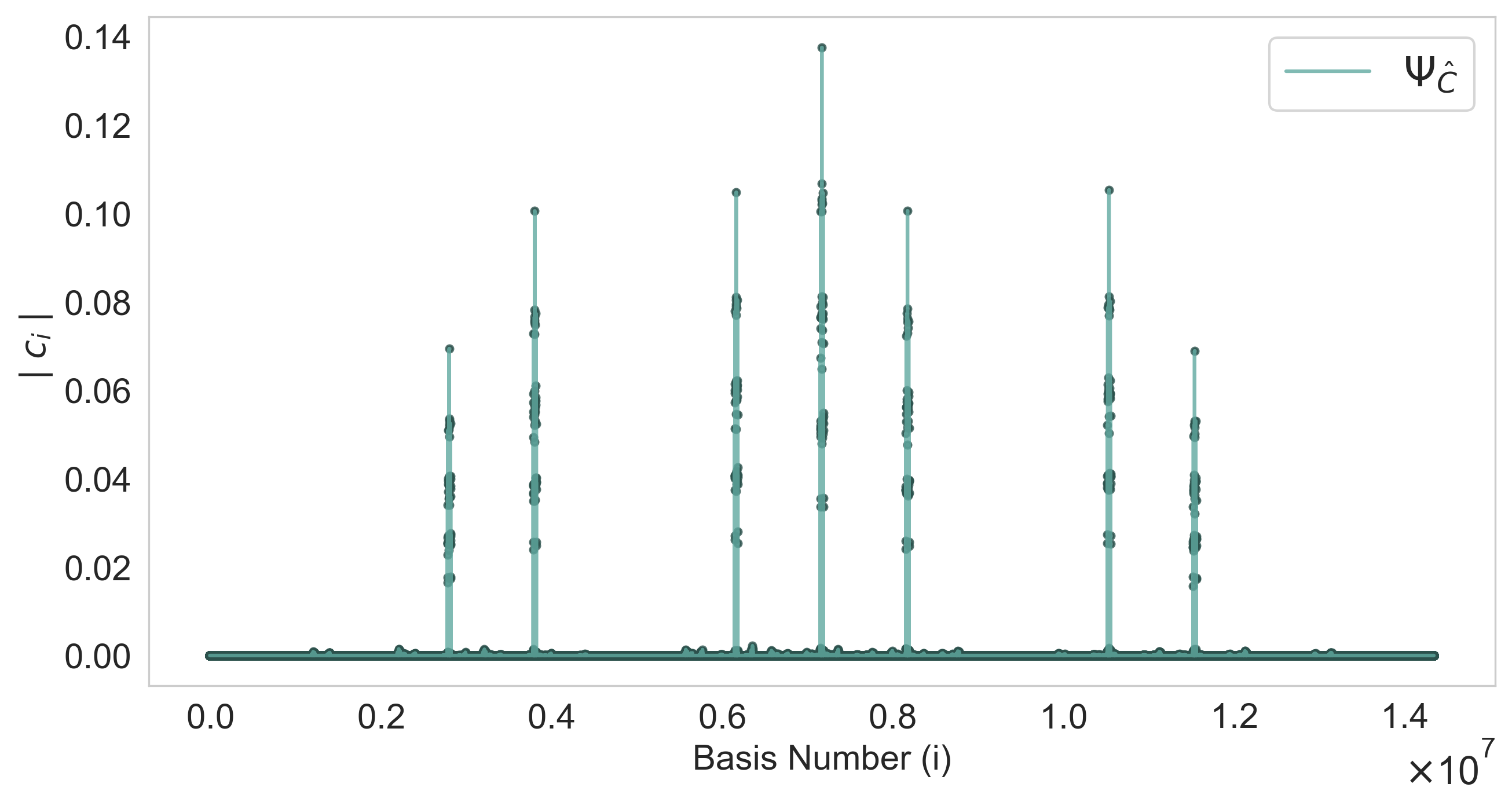}
    \end{subfigure}
    \caption{The states $\Psi_{\hat{C}_{\mathrm{TRC}}}$ and $\Psi_{\hat{C}}$ which were obtained as a result solving the $\hat{C}_{\mathrm{TRC}}$ and $\hat{C}$ constraints respectively in a $\jmax = 1$ cutoff are shown. The states belong to Hilbert spaces of the dimensions $\approx 14\times 10^6$. On the abscissa, the label of the basis state is shown while on the ordinate, the corresponding amplitude. We notice that the zero charge state $(\Vec{0}, \Vec{0}, \Vec{0}, \Vec{0}, \Vec{0})$ is the most contributing basis state in both solutions.}
    \label{fig:compareConstraintStates}
\end{figure}

\begin{figure}[h]
    \centering
    \begin{subfigure}{.5\textwidth}
        \centering
        \includegraphics[width=1.1\linewidth]{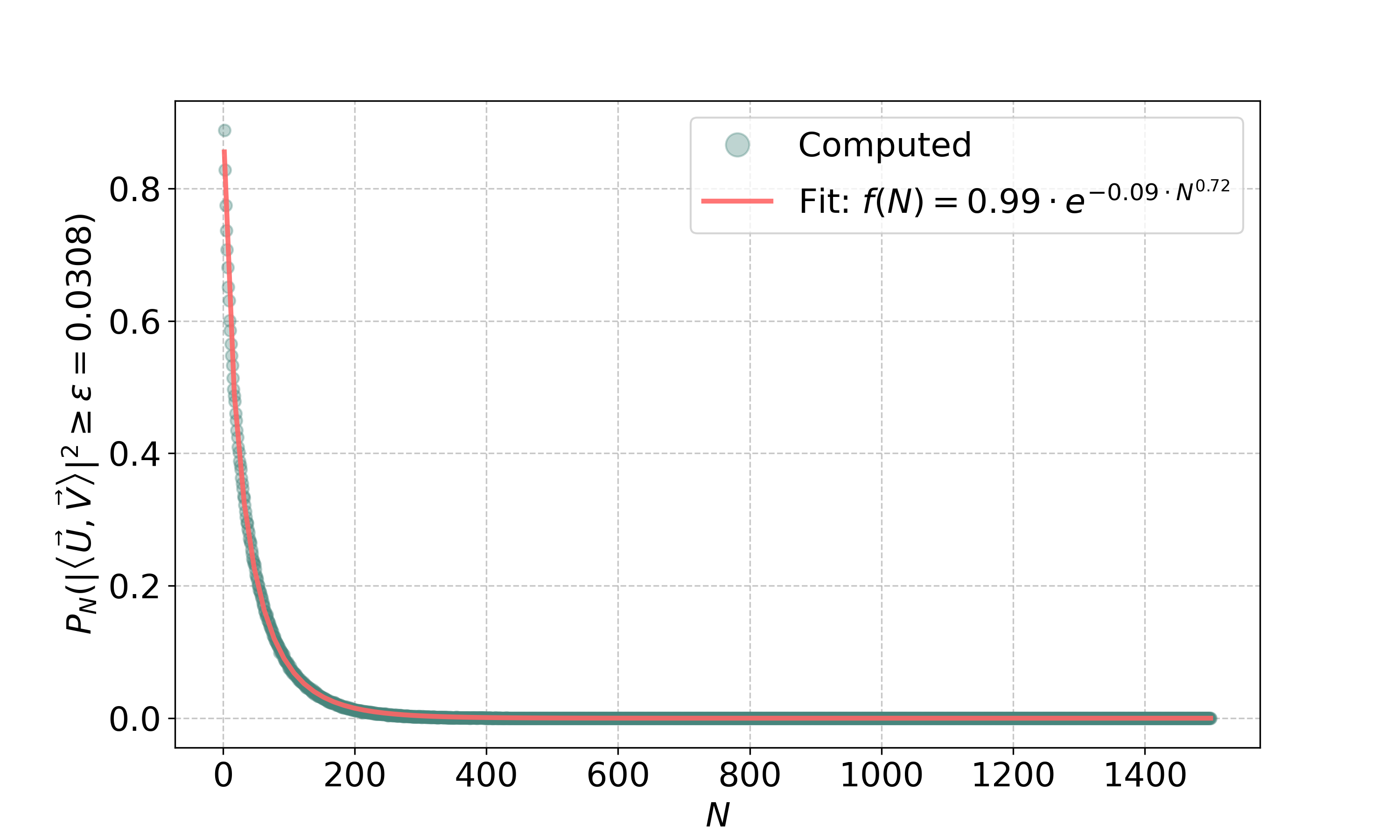}
    \end{subfigure}%
    \begin{subfigure}{.5\textwidth}
        \centering
        \includegraphics[width=1.1\linewidth]{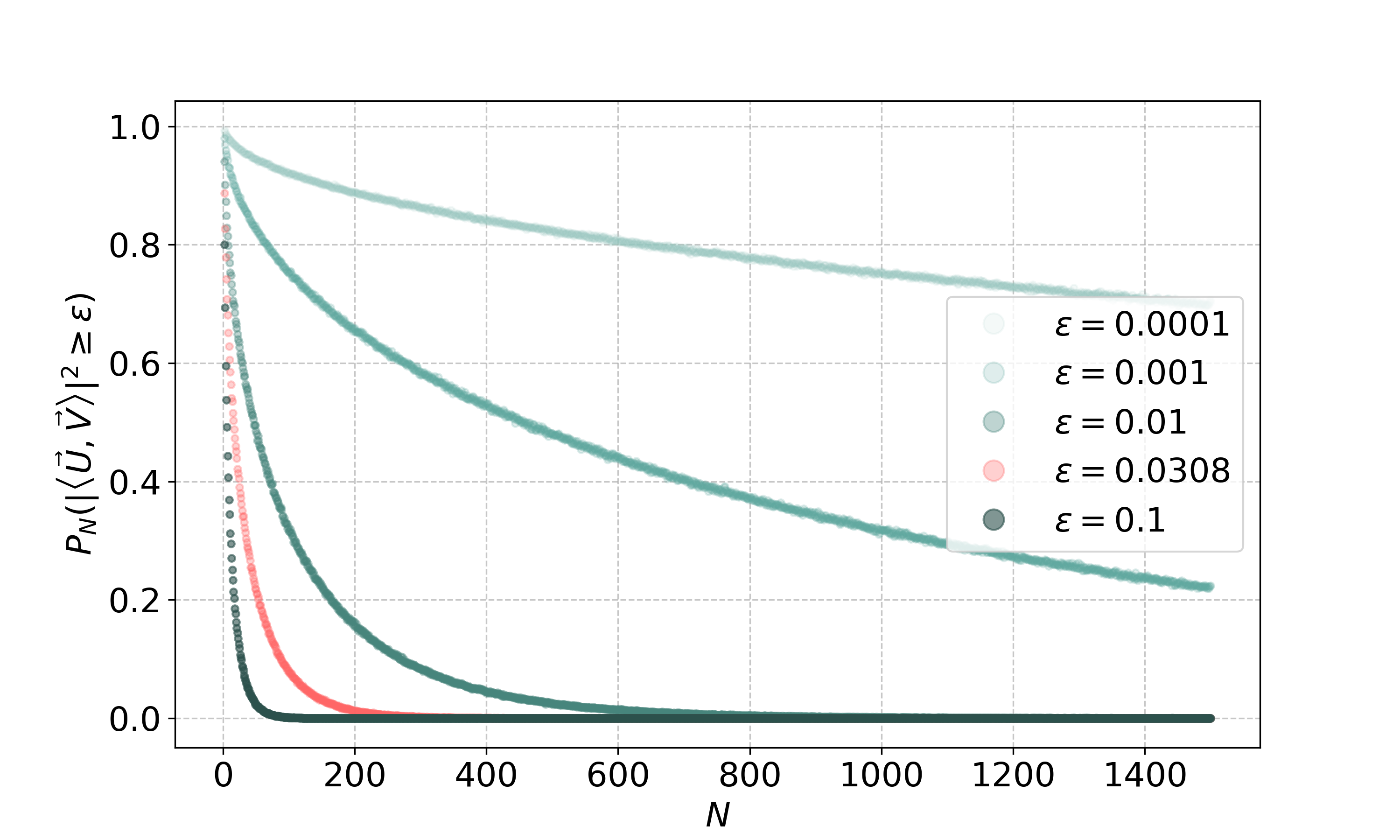}
    \end{subfigure}
    \caption{On the left, $P(|\langle \Vec{U}, \Vec{V}\rangle|^2 \geq \epsilon = 0.0308)$ of two randomly chosen normalised unit vectors $\Vec{U}, \Vec{V}$ on $S^{N-1} \subset \mathbb{R}^N$ is shown in green and an exponential fit function is shown in red. On the right, the same probability is shown for different values of $\varepsilon$.}
    \label{fig:probabilityEstimates}
\end{figure}

At this point, the NNQS ansatz was used to solve both the $\hat{C}$ and $\hat{C}_{\mathrm{TRC}}$ constraints, arriving in the true solution space of the former and the kernel of the latter. It was observed qualitatively that these two spaces have little overlap. Now, we quantify this overlap by investigating the contributing basis states of both the solutions.
\newline
For the $\jmax = 1$ cutoff used, one can obtain the $\sim 10^6$ amplitudes for each of the variational states. The first point would be to visualise the two states to determine the type of contributing basis states as done in Figure \ref{fig:compareConstraintStates}.
\newline
Figure \ref{fig:compareConstraintStates} shows the solution $\Psi_{\hat{C}}$ of the $\hat{C}$ constraint on the right and the state $\Psi_{\hat{C}_{\mathrm{TRC}}}$ near the kernel of $\hat{C}_{\mathrm{TRC}}$ that the NNQS ansatz arrived at on the left. Both simulations were conducted in a $\jmax = 1$ cutoff. In both, it is observed that the most contributing basis state happens to be the state with all charge vectors being zero. However, it is immediate to see that otherwise the two states do not have much in common. The mean of the absolute value of the amplitudes of $\Psi_{\hat{C}}$ was computed to be $1.345 \times 10^{-6}$ while for $\Psi_{\hat{C}_{\mathrm{TRC}}}$ it was $1.194\times 10^{-7}$. This shows that the solution of $\Psi_{\hat{C}}$ is peaked more strongly, which is also visually confirmed in Figure \ref{fig:compareConstraintStates}.
\newline
To conduct a more than qualitative comparison, we begin by investigating the contributing basis states of both $\Psi_{\hat{C}}$ and $\Psi_{\hat{C}_{\mathrm{TRC}}}$. We will denote by \textit{contributing basis states} ones which have an amplitude which falls above or equal to a certain threshold. This threshold is taken to be $\epsilon = 1.0\times 10^{-6}$. Furthermore, to quantify how much do the amplitudes of these contributing basis states actually contribute to their respective variational states, we compute the \textit{significance} of these states by computing the value
\begin{equation}
\label{eq:significance}
    ||\Psi^\epsilon|| = \sqrt{\sum_{I_\epsilon} (c_{I_\epsilon})^2},
\end{equation}
where we sum over the amplitudes of the basis states numbered by $I_\epsilon$ such that $c_I > \epsilon$.
Lastly, we narrow the criteria even further to determine the similarity between the two states. Therefore, we consider the \textit{common} contributing basis states: basis states which contribute above the chosen cutoff $\epsilon$ for \textit{both} $\Psi_{\hat{C}}$ and $\Psi_{\hat{C}_{\mathrm{TRC}}}$. The significance of these common contributing basis states can be computed also using equation \eqref{eq:significance} above while now we only keep in account the amplitudes of those common states. The results obtained are shown in Table \ref{tab:quantum_state_analysis}
\begin{table}[htbp]
    \centering
    {\renewcommand{\arraystretch}{1.5}
    \begin{tabular}{c|c|c|c|c}
        \rowcolor{lightergreen!50}
        \multirow{2}{*}{\cellcolor{lightergreen!50}State} & \multicolumn{2}{c|}{Contributing States} & \multicolumn{2}{c}{Common Contributing States} \\
        \cline{2-5} 
        \rowcolor{lightergreen!50}
        \multirow{-2}{*}{\cellcolor{lightergreen!50}State}
         & \textit{Amount} & \textit{Significance} & \textit{Amount} & \textit{Significance} \\
        \hline
        $\Psi_{\hat{C}}$ & 39737 & 0.999999995354185 & \multirow{2}{*}{1299} & 0.4815059317489464 \\
        $\Psi_{\hat{C}_{\mathrm{TRC}}}$ & 2226 & 0.9999999996964684 & & 0.9999993127698954 \\
    \end{tabular}
    }
    \caption{The amount and significance of the basis states with amplitudes $\geq \epsilon = 1.0 \times 10^{-6}$ are shown in the Contributing States column for both  $\Psi_{\hat{C}}$ and $\Psi_{\hat{C}_{\mathrm{TRC}}}$. The Common Contributing States column shows the amount and significance of the contributing basis states which are common in both  $\Psi_{\hat{C}}$ and $\Psi_{\hat{C}_{\mathrm{TRC}}}$. Here, $\dim\hilb_{\gamma} \approx 14 \times 10^6$.}
    \label{tab:quantum_state_analysis}
\end{table}
\newline
As seen from Table \ref{tab:quantum_state_analysis}, the amount of contributing basis states with amplitudes above or equal to the chosen cutoff $\epsilon = 1.0 \times 10^{-6}$ for both $\Psi_{\hat{C}}$ and $\Psi_{\hat{C}_{\mathrm{TRC}}}$ is very low, constituting roughly 0.2769\% and 0.01551\% respectively of the Hilbert space (which at this cutoff has dimensions $\dim\hilb_{\gamma} \approx 14 \times 10^6$). Nevertheless, these relatively few, and especially so in the case of $\Psi_{\hat{C}_{\mathrm{TRC}}}$, basis states have amplitudes which contribute \textit{very} strongly to their respective states. Furthermore, if one narrows the focus only on common contributing states, one finds only 1299 states (0.00905\% of $\dim\hilb_{\gamma}$) which both contribute with a value $\geq \epsilon$ and are common between the two states. Yet, these 1299 states contribute relatively strongly although not to a similar degree to each respective state. We note however that all of these values are much more larger than the dimensions of the gauge invariant subspace which in this case is $\dim\hilb_{\gamma}^G = ((2\jmax + 1)^2)^3 = 729$.
\newline
We conduct the last measure of comparison by computing the inner product of the two states $\lvert\langle \Psi_{\hat{C}_{\mathrm{TRC}}} \mid \Psi_{\hat{C}} \rangle\rvert^2$ as well as calculating the angle between them. In doing so, it is seen that
\begin{equation}
    |\langle \Psi_{\hat{C}_{\mathrm{TRC}}} \mid \Psi_{\hat{C}} \rangle|^2 \approx 0.0308 \quad , \quad \arccos|\langle \Psi_{\hat{C}_{\mathrm{TRC}}} \mid \Psi_{\hat{C}} \rangle| \approx 1.394\mathrm{ rad}.
\end{equation} 
At first sight, these results seem may seem to be a little discouraging, suggesting that the solution of $\hat{C}$ and this state near the kernel of $\hat{C}_{\mathrm{TRC}}$ are unrelated. However, they are not exactly orthogonal to one another either. As such, we now quantify the significance of this similarity. 
\newline
To do this, one can think of random state picking in this Hilbert space. This would give a measure of significance by asking the question: 
what is the probability that two randomly chosen states from the Hilbert space have an overlap larger or equal to that of $\Psi_{\hat{C}}$ and $\Psi_{\hat{C}_{\mathrm{TRC}}}$? If one considers the entire Hilbert space $\hilb_{\gamma}$, which contains states with complex valued coefficients, then this question can be addressed exactly by random picking from the probability $N$-simplex (see Appendix \ref{app:C}) where $N = \dim_{\mathbb{C}}\hilb_{\gamma}$. In this case, one sees that the probability that two states have the overlap of $0.0308$ obtained above is
\begin{equation}
    P_N^{\mathbb{C}}(|\langle \Psi_{\hat{C}_{\mathrm{TRC}}} \mid \Psi_{\hat{C}} \rangle|^2 \geq 0.0308) \sim 10^{-194953},
\end{equation}
where $N := \dim_{\mathbb{C}}\hilb_{\gamma} \sim 10^6$. While this looks impressive at first sight, it should of course be expected that both states are much closer than random due to the fact that they are both approximately gauge invariant. 
If one considers only the gauge invariant sector, then $N^G := \dim_{\mathbb{C}}\hilb_{\gamma}^G = ((2\jmax + 1)^2)^3 = 729$ for $\jmax = 1$. Repeating the same calculation shows that this would again yield a negligible probability
\begin{equation}
    P_{N^G}^{\mathbb{C}}(|\langle \Psi_{\hat{C}_{\mathrm{TRC}}} \mid \Psi_{\hat{C}} \rangle|^2 \geq 0.0308) \sim 10^{-10}.
\end{equation}
In this work we explored states with real valued coefficients in the computational basis (to further downsize the computational requirements, also see Appendix C of \cite{Sahlmann:2024pba}). A precise statement regarding the probability of two states being $\varepsilon$ similar therefore needs to take that into consideration. The analogous picking process restricted to only the subspace $\hilb^{\mathbb{R}}_{\Tilde{\gamma}}$ composed only of states with real valued coefficients amounts to random picking of uniformly distributed vectors on the unit sphere $S^{N - 1} \subset \mathbb{R}^{N}$ where $N = \dim_{\mathbb{R}}\hilb^{\mathbb{R}}_{\Tilde{\gamma}}$. An exact analytical expression for $P_N^{\mathbb{R}}(|\langle \Psi_{\hat{C}_{\mathrm{TRC}}} \mid \Psi_{\hat{C}} \rangle|^2 \geq \varepsilon)$ was not obtained in this case. However, numerical estimates are shown in Figure \ref{fig:probabilityEstimates}.
\newline
In Figure \ref{fig:probabilityEstimates} on the left, a numerical estimate of the probability that the inner product squared of two randomly chosen normalised unit vectors $\Vec{U}, \Vec{V}$ which lie on the unit sphere $S^{N-1} \subset \mathbb{R}^N$ (for different $N \in [2, 1500]$) is $\geq \varepsilon = 0.0308$. Here, for every $N$, two normalised unit vectors which lie on the hypersphere were chosen at random. The absolute value squared of their inner product was computed and stored if it had a value $\geq \varepsilon$. This process was repeated for $10^7$ trials for every $N$, each time generating new vectors and repeating the computation. The probability, for a given $N$, of the inner product squared of such random vectors being $\geq \varepsilon$ was then computed as the number of such occurrences divided by the number of trials conducted. The results were then fitted with an exponential fit function as shown in the figure as well. On the right in Figure \ref{fig:probabilityEstimates}, the same simulation is done for different values of $\varepsilon$ with a number of trials being $10^6$. The results indicate that for $n \rightarrow \infty$, random vectors on $S^{N-1}$ are very likely to be almost orthogonal as the likelihood of them being $\varepsilon$ similar decreases for large $N$, a result also known from concentration of measures on the sphere \cite{Berestycki2019ConcentrationOM}. 
\newline
Given the exponential decay shown in the figure above, we can then address the random picking procedure on $\hilb_{\gamma}^{\mathbb{R}}$ as well. We see that even if one restricts to the real subspace, then 
\begin{equation}
\label{eq:probReal}
    P_N^{\mathbb{R}}(|\langle \Psi_{\hat{C}_{\mathrm{TRC}}} \mid \Psi_{\hat{C}} \rangle|^2 \geq 0.0308) \sim 10^{-5559}.
\end{equation}
Further, if one restricts to the gauge invariant sector of the real subspace, the probability is still very small, with a value of
\begin{equation}
\label{eq:probRealGauss}
    P_{N^G}^{\mathbb{R}}(|\langle \Psi_{\hat{C}_{\mathrm{TRC}}} \mid \Psi_{\hat{C}} \rangle|^2 \geq 0.0308) \sim 10^{-5}.
\end{equation}
We do note, however, that any numerical estimates are bound to errors. For example, as can be seen from Figure \ref{fig:probabilityEstimates} on the left, the probabilities obtained are zero if one considers large $N$. While this may be due to numerical precision limits, it may also be due to choosing a number of trials which may not be sufficient. What concerns us in this work is $N = 729$\footnote{since the dimensions of the gauge invariant subspace of the real subspace is $\dim_{\mathbb{R}}\hilb_{\gamma}^{\mathbb{R}, G} = 729$}. As such, two sets of 1000 simulations were done specifically for such choice of $N$. In the first set, each simulation included $10^6$ trials and in the second, $10^7$ trials. In none of all 2000 simulations was any occurrence observed. However, the results of these simulations showed that for some dimensions before and after $N = 729$, a probability of $10^{-6}$ was observed. Hence, it is fair to conclude that for $N = 729$, a probability of $10^{-6}$ is also expected. This is not far off from what is obtained through the fit function as shown in equation \eqref{eq:probRealGauss}. We also note that the value of the exponential decay in the fit function shown in Figure \ref{fig:probabilityEstimates} \emph{did not} change when considering only the non-zero values ($N \leq 200$). Thus, the values which are identically zero, irrespective of the reason, do not affect the resulting fit function. In all cases, it is expected to follow an exponential decay as can be seen from the concentration of the measure on the sphere \cite{Berestycki2019ConcentrationOM}.
\newline
To conclude, the remarkable picture is that if picked at random, whether one looks at the entire Hilbert space or the real subspace only, and whether one restricts to investigating the gauge invariant sector or not, there is in practice zero probability of the two states having even this small overlap if picked at random. The take away here is that while the two regularisations produce constraints which look radically different, and the solution of the master constraint $\hat{C}$ and the states near the kernel of the constraint $\hat{C}_{\mathrm{TRC}}$ may look qualitatively very different, the numerical findings suggest that the two quantisations may have something in common as this significant and non-trivial overlap in the states obtained when solving them could not have occurred randomly.
\newline
We now make a few cautionary comments regarding the findings: 
\begin{enumerate}
    \item All of the results obtained in this study regarding the TRC were conducted in a $\jmax = 1$ cutoff due to computational limitations. Therefore, it remains unknown how these results translate for higher $\jmax$ cutoffs. We have observed that the solutions of the master constraint $\hat{C}$ changed under different cutoffs (e.g. number of contributing basis states, etc.), and as such it is not completely excluded that the same would happen for $\hat{C}_{\mathrm{TRC}}$. 

    \item The results shown in this section correspond to \textit{one} state near the kernel of $\hat{C}_{\mathrm{TRC}}$. During simulations, five different states near the kernel were obtained including the one presented in this work. These states looked qualitatively and quantitatively different. That is, the contributing states were different to the ones shown in this work (Figure \ref{fig:compareConstraintStates}) and the contributing and common contributing amplitudes were also different. In such simulations, the inner product $\mid\langle \Psi_{\hat{C}_{\mathrm{TRC}}} \mid \Psi_{\hat{C}} \rangle\mid^2$ had different values ranging from approximately $0.015$ to $0.0012$. Nevertheless, conducting random picking of states in $\hilb_{\gamma}^{\mathbb{R}}$ of such large dimensions shows that such results are however still significant as the probability $P_N^{\mathbb{R}}(|\langle \Psi_{\hat{C}_{\mathrm{TRC}}} \mid \Psi_{\hat{C}} \rangle|^2 \geq \varepsilon)$ of randomly picking the state $\Psi_{\hat{C}}$ starting from the state $\Psi_{\hat{C}_{\mathrm{TRC}}}$ with $\varepsilon \sim 0.001 , \dots , 0.01$ and $N = \dim_{\mathbb{R}}\hilb_{\gamma}^{\mathbb{R}} \sim 10^{6}$ is still zero (in the order of $10^{-271}$ to $10^{-2913}$ respectively). 

    \item Computing $\expect{\tr \hat{h}_{\alpha_k}}$ in the state $\Psi_{\hat{C}_{\mathrm{TRC}}}$ yields different values compared to $\Psi_{\hat{C}}$. Specifically, the values are far less than 3, which hint that the obtained state is not flat. This is, however, to be expected as one can can write the state $\Psi_{\hat{C}_{\mathrm{TRC}}}$ as
    \begin{equation}
        \Psi_{\hat{C}_{\mathrm{TRC}}} = \Psi_{\perp} + \lambda \Psi_{\hat{C}},
    \end{equation}
    where $\Psi_{\perp}$ represents states orthogonal to $\Psi_{\hat{C}_{\mathrm{TRC}}}$ and $\lambda$ a constant proportional to $\varepsilon$. The expectation value $\expect{\tr \hat{h}_{\alpha_k}}$ will therefore include contributions of the form
    \begin{equation}
        \expect{\tr \hat{h}_{\alpha_k}}_{\Psi_{\hat{C}_{\mathrm{TRC}}}} = \expect{\tr \hat{h}_{\alpha_k}}_{\Psi_{\perp}} + |\lambda|^2\expect{\tr \hat{h}_{\alpha_k}}_{\Psi_{\hat{C}}} + \lambda\langle \Psi_{\perp}, \hat{h}_{\alpha_k} \Psi_{\hat{C}}\rangle + \overline{\lambda}\langle \Psi_{\hat{C}}, \hat{h}_{\alpha_k} \Psi_{\perp}\rangle.
    \end{equation}
    Of all such terms, the contributions purely from $\Psi_{\hat{C}}$ have the smallest value. As such, it is anticipated that such expectation values would differ in their values when computed in $\Psi_{\hat{C}}$ as opposed to $\Psi_{\hat{C}_{\mathrm{TRC}}}$.
    \item We should remind the reader that we have fixed the lapse function, $N(v)=1$. Thus the condition 
\begin{equation}
\expect{\hat{C}_{\mathrm{TRC}}}_\Psi = \text{min}
\end{equation}
is only one of many necessary conditions. 

\item So far we have not discussed the diffeomorphism constraint. Note that it is part of the curvature constraint and hence included in $\hat{C}$ but it is not included in $\hat{C}_{\mathrm{TRC}}$, however. Thus it would have to be implemented separately when implementing $\hat{C}_{\mathrm{TRC}}$. From \eqref{eq:solutionnnn} one can deduce that implementing the diffeomorphism constraint in the current situation would only require a symmetrisation of the states with respect to the graph symmetries GS$_\gamma$ of $\gamma$. For the $\gamma$ considered in this work, $\text{GS}_\gamma \equiv S_2$ and the non-trivial graph symmetry acts via 
\begin{equation}
    (\vec{m}_1, \ldots ,\vec{m}_5)\longmapsto (-\vec{m}_5, -\vec{m}_2, -\vec{m}_4, -\vec{m}_3, -\vec{m}_1)
\end{equation}
on the computational basis. We have \emph{not} implemented the corresponding averaging in the present work.
\end{enumerate}

\subsection{Possible improvements}
\label{sec:betterArchitecture}
In this section, we briefly comment on the technical/computational issues encountered in solving the $\Uqone^3$ model. In principle, there are two roads one could take in attempting to solve this model: (i) exploring the entire Hilbert space for solutions or (ii) exploring only the gauge invariant subspace. If one follows (ii), then since $\dim\hilb_{\gamma}^{G} \ll \dim\hilb_{\gamma}$, this problem no longer exists. However, to do so, one would be required to implement networks which are inherently gauge invariant thus identically satisfying the Gau{\ss} constraint. In such an approach, it is also expected to arrive at solutions more efficiently both in terms of computational resources and runtime as the space required to be explored is rather small which in turn allows for exploring complicated operators at high cutoff values. This approach will be discussed in future work. In this work, we follow (i), and thus are faced with the problem of large $\dim\hilb_{\gamma}$. In what follows, some avenues to explore to better the situation of the (i) approach are presented. 
\newline
First, the current network architecture can be more finely tuned to such large dimensional spaces, as the one used in this work was primarily implemented for the comparatively smaller $\Uqone$ model. Put simply, one has now a problem where the number of input neurons of the network is large. This is because of having three copies of the same graph to accommodate for the charge vectors on the edges. This becomes more pronounced if one goes further beyond the simple graph considered in this work. One work around is to consider networks which specifically include dimensionality reduction. For example, a U-Net \cite{10.1007/978-3-319-24574-4_28}, typically used in image segmentation problems, contains an up-sampling and a down-sampling branch to deal with this issue. 
\newline
The suggested network redesign addresses the issue of large graphs. However, even for the smallest possible graph considered in this work, one eventually hits another hurdle which is the number of allowed states per dual vertex, that is $\dim\hilb_{\gamma}$ growing quickly for small $\jmax$ increments. This issue is realised in the MCMC process. The Metropolis type samplers used in this work are either local or 2-local (see Appendix B in \cite{Sahlmann:2024pba}). Such samplers do not efficiently explore the Hilbert space. More sophisticated samplers would be required to remedy this problem. For example, one could introduce Metropolis type samplers which sample only gauge invariant configurations. In this manner, one would not have a gauge invariant network but a gauge respecting sampler, effectively also exploring only gauge invariant states. 
\newline
In all cases, these issues remain present for complicated gauge groups and therefore should be addressed appropriately. This is especially so if one wishes to construct a universal network capable of solving any quantum gravity models. At the moment, this is neither completely necessary nor known to be feasible. Further, it is for now unclear how this translates if one considers the SU(2) gauge instead of the simpler $\Uone^3$ group. Naturally, one expects to have similar issues. Thankfully, for this gauge group, whether one deals with 2+1 or 3+1 spacetime dimensions, one can always solve the simple $\Uqone$ model to verify the results obtained in the higher gauge dimensional models enabling one to continue this endeavour.

\section{Discussion and outlook}
In this work, we expanded upon the work done in \cite{Sahlmann:2024pba} where we have applied the NNQS ansatz to solve two constraints of LQG quantised 3d Euclidean gravity in the weak coupling limit and explored quantum geometric observables of LQG. We started by considering the classical theory, in which it was shown that Euclidean gravity in 3-dimensions in the weak coupling limit was a 3-dimensional BF-theory with a $\Uone^3$ Lie group. The classical theory was subject to the two rather simple constraints
\begin{equation}
    F^{I}_{ab} := 2\partial_{[a}\Tilde{\omega}^{I}_{b]} = 0 \quad , \quad \partial_{a} e^{a}_{I} = 0.
\end{equation}
The first of the constraints is the curvature constraint, ensuring flatness of the connection while the second is the Gau{\ss} constraint generating gauge transformations. The theory was quantised using LQG methods. To facilitate the computational implementation, we truncated the kinematical degrees of freedom by (a) fixing a graph and (b) limiting the values the charge vectors are allowed to have in the model. This resulted in what we called a $\Uqone^3$ BF-theory. Upon quantisation, we obtained quantum analogs of the two classical constraints where the curvature ensured flatness of minimal loop holonomies and the Gau{\ss} constraint ensured charge vector conservation at the vertices of the graph. We formulated a master constraint $\hat{C}$ consisting of the squares of these two constraints. Further, we implemented a quantum Hamilton constraint $\hat{H}$ following the strategy proposed in \cite{Thiemann:1997ru} after which we constructed the Hermitian constraint $\hat{C}_{\mathrm{TRC}} = \hat{H} + \hat{H}^\dagger + \hat{G}$.
\newline
The NNQS ansatz was employed to solve the master constraint $\hat{C}$ of this quantum theory using a network with the same architecture as used in previous work \cite{Sahlmann:2024pba}. The architecture proved successful, where it obtained high accuracy for relatively low charge cutoff values. Due to computational limits, such a model was not possible to solve using exact numerical methods. Hence, this demonstrated the ability of the NNQS ansatz to explore physical systems which would otherwise not be solvable at least using traditional numerical techniques. The quantum fluctuations of the minimal loop holonomy operators were computed in which it was seen that with higher charge cutoff, one starts to recover the continuum theory and obtain truly flat and gauge invariant solutions.
\newline
The last part of the work concerned exploring the some operators of 3d LQG. Specifically, we started by numerically verifying the results found in the literature by showing that the 3d quantum volume operator of LQG indeed does not necessarily vanish on 2 or 3-valent vertices as the 4d volume operator of LQG would. Next, we considered the constraint $\hat{C}_{\mathrm{TRC}}$. In such a case, the operator was only investigated in the smallest possible cutoff in which $\jmax = 1$ due to technical and computational limitations which deemed that necessary. Nevertheless, it was seen that the NNQS ansatz was able to arrive at states near the kernel of $\hat{C}_{\mathrm{TRC}}$. It was observed that in this specific state obtained in this work, the volume operator behaved in a qualitatively similar manner to the behaviour seen for states in the solution space of $\hat{C}$. The solution space of $\hat{C}$ and the states near the kernel of $\hat{C}_{\mathrm{TRC}}$ were compared. We showed that both the states have common basis states which contribute relatively strongly, but not equally. The inner product of the two states was shown to be of significance despite its small value. This indicates that while the constraints differ radically in their regularisation, the two states had a significant non-trivial similarity between them. 
\newline
As this work pushes further the application of NNQS in canonical quantum gravity, we have to note not only the results but also the shortcomings and issues encountered. It was observed that the network's architecture used is sub-optimal, leaving room for exploring architectures better suited for very large Hilbert spaces and very large graphs. Furthermore, this issue may become more evident if one considers more complicated gauge groups. Hence, it is suggested that one considered inherently gauge invariant implementations of network to be used for the ansatz. Furthermore, the Metropolis-Hastings sampler can also be tailored to explore very large Hilbert spaces more efficiently. 
\newline
Nevertheless, it is now shown that NNQS can be used to solve simplified LQG models of pure gravity in 3-dimensions. Although the model considered here consisted of only pure 3d Euclidean gravity in the weak coupling limit, one can in principle consider more interesting models. This can range from considering different topologies for $\Sigma$ such as a torus universe to the inclusion of matter or scalar fields. Now with the ability to solve the quantum Hamilton constraint of physical models, one can also investigate interesting questions such as the correlation range and entanglement of the solution. Lastly, one can expand upon this by considering now 4d gravity in the weak coupling limit. These avenues will be explored in upcoming work. 

\ack
H.S. acknowledges the contribution of the COST Action CA18108. The authors gratefully acknowledge the scientific support and HPC resources provided by the Erlangen National High Performance Computing Center (NHR@FAU) of the Friedrich-Alexander-Universität Erlangen-Nürnberg (FAU). The hardware is funded by the German Research Foundation (DFG). The numerical simulations were carried out using Netket \cite{CARLEO2019100311}.

\clearpage

\section*{Appendices}
\renewcommand{\thesubsection}{\Alph{subsection}} 
\setcounter{footnote}{0} 

\subsection{Computational requirements and limits}
\label{app:B}
\begin{figure}[h]
    \centering
    \begin{subfigure}{.5\textwidth}
        \centering
        \includegraphics[width=1.0\linewidth]{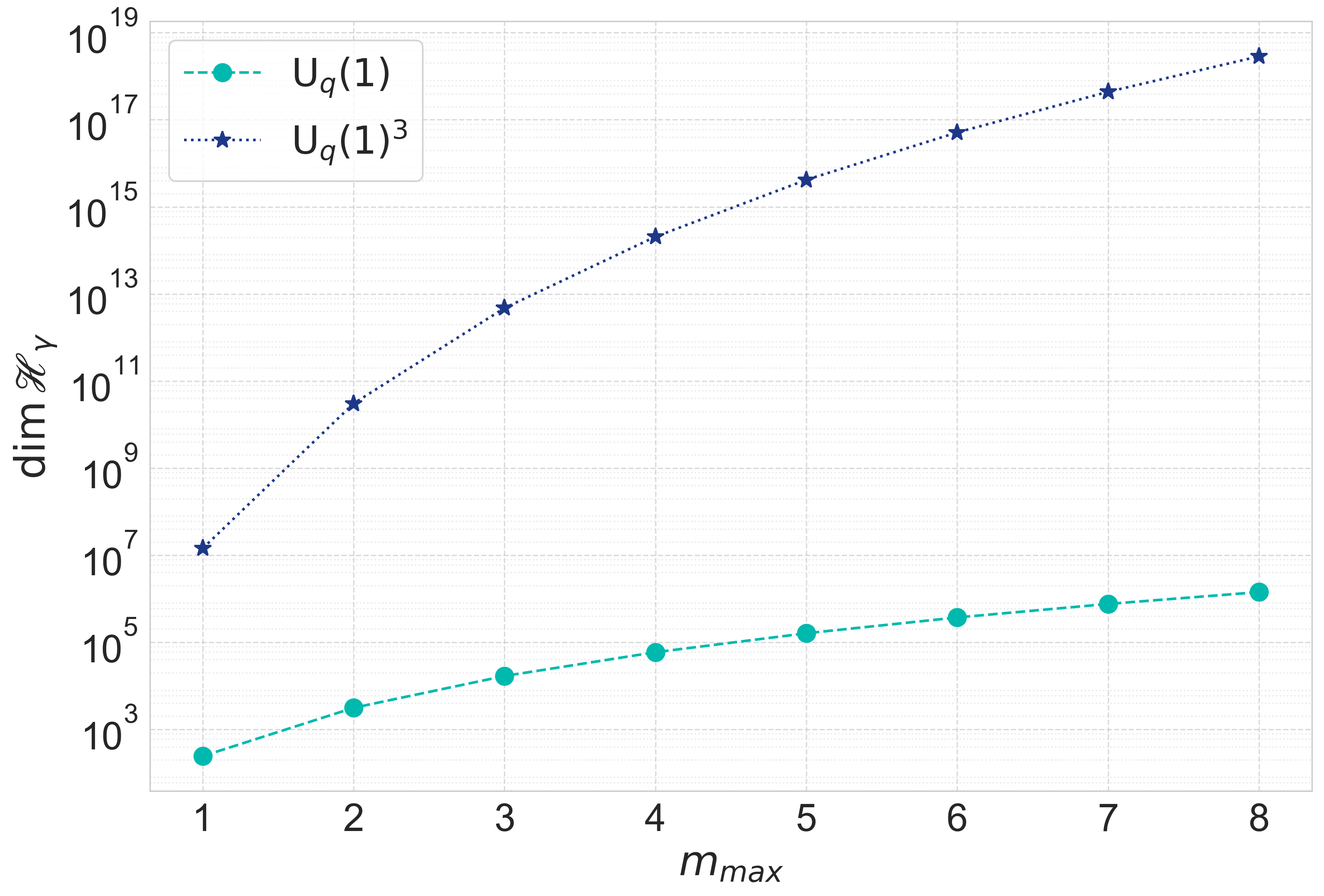}
        \label{fig:comparingMemory_sub1}
    \end{subfigure}%
    \begin{subfigure}{.5\textwidth}
        \centering
        \includegraphics[width=1.0\linewidth]{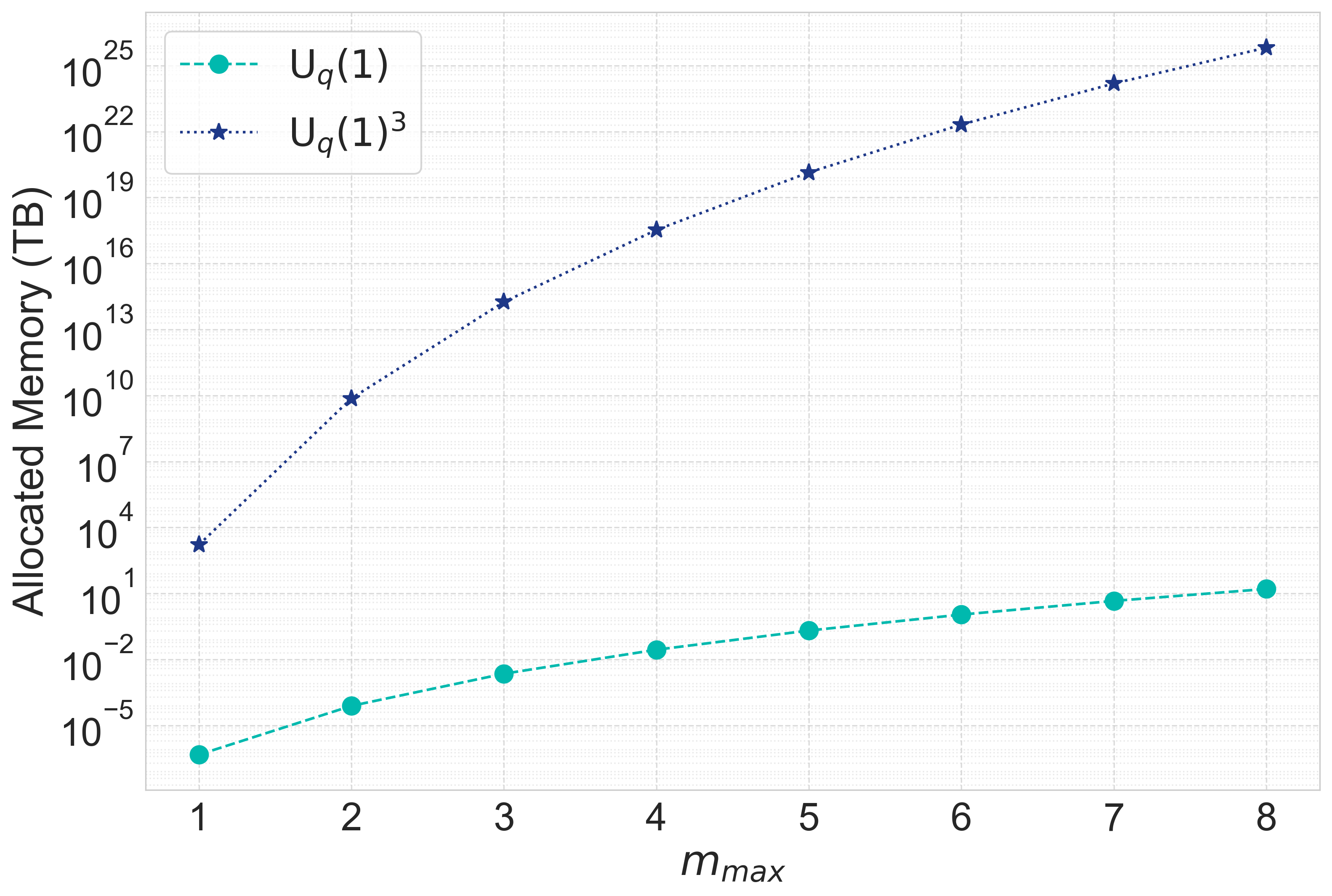}
        \label{fig:comparingMemory_sub2}
    \end{subfigure}
    \caption{Two log plots are shown. On the left, the dimension of the Hilbert space in both the $\Uqone$ and $\Uqone^3$ BF-theory models are shown while on the right is the required RAM which is to be allocated if one wishes to do exact diagonalisation for the constraint in each of the models at different $\jmax$ cutoffs.}
    \label{fig:comparingMemory}
\end{figure}

The dimensions of the Hilbert space of the $\Uqone^3$ model is $D := \dim\hilb_{\gamma} = (D^{(1)})^3$ where by $D^{(1)} := (2\jmax + 1)^N$ we denote the dimensions of the Hilbert space of a $\Uqone$ model considered in \cite{Sahlmann:2024pba} over an $N$ vertex dual graph. 
\newline
We now give an estimate and comparison on the computational resources required for both models. In this comparison, we let $N = 5$ and $\jmax = 1$. Our operators in the computational framework are composed of real numbers which are stored in a 64-bit floating point datatype on the computer, requiring 8 bytes in memory\footnote{assuming the double (\texttt{float64}) datatype in C, as done in our work, according to the IEEE 754-2008 Standard for Binary Floating-Point Arithmetic}. If one wishes to store the amplitudes of the variational state (arrays of size $\dim\hilb_{\gamma}$), then one needs 1.944 Kilo Bytes for the $\Uqone$ model, and 144.8 Mega Bytes (MB) for the $\Uqone^3$ model. If one wishes to solve the models using exact numerical methods, this would require (typically) a sparse representation of the matrix representing the constraint. For the $\Uqone^3$ case, where $D \sim 10^6$ for $\jmax = 1$, then one needs to allocate 1.647 Peta Bytes (PB) (1647000 Giga Bytes (GB)) for the $D\times D$ matrix representing the constraint. Figure \ref{fig:comparingMemory} shows the log plots of the memory requirements for exactly solving the constraint and the dimensions of the Hilbert spaces for both models.
\newline
The memory requirement for exact diagonalisation as well as the dimensions of the Hilbert spaces of both $\Uqone$ and $\Uqone^3$ models are shown in Figure \ref{fig:comparingMemory} on the right and left respectively. This required memory is to be allocated as RAM, as the memory is allocated during runtime. For the $\jmax = 8$ cutoff, one would need $65.548 \times 10^{6}$ Quetta Bytes (QB) for storing the sparse matrix of the $\Uqone^3$ constraint, where 1 QB =  $10^{18}$ TB = $10^{21}$ GB. Clearly, these numbers fall beyond the possible computational limits even though this comparison was done for the smallest possible non-trivial graph. 
\newline
The NNQS ansatz parameterises the wave-function in terms of a neural network and hence, the number of parameters required to be determined to fully characterise the wave-function are the number of weights connecting the neurons in the network. These parameters are often far fewer than $\dim\hilb$. Further, modern computational libraries and methods allow for automatic differentiation, accelerated linear algebra computations and just-in-time compilation all of which reduce the demand on computational resources. In the current implementation, one does not even need a sparse matrix representation of the constraint, rather only the non-zero matrix elements thus reducing the memory demand in the simulations as well.

\subsection{Series expansion of the $\hat{V}_{v}^2$ operator}
\label{app:A}
Consider the operator of the form
\begin{equation}
    \hat{V}_{v}^2 := \sum_I\left( \sum_{e, e' \text{ at } v} \sign(e, e') \epsilon_{IJK} X^{J}_{e} X^{K}_{e'}\right)^2.
\end{equation}
The aim is to Taylor expand the above operator. Let us define $f(x) = x$ where $x$ is the entire right hand side of the above equation. In the following, we first show the general closed form of the series expansion of $g(x) = \sqrt{f(x)} = \sqrt{x}$ and following that, for $h(x) = \sqrt{g(x)} = \sqrt{\sqrt{x}}$ as well.
\newline
The Taylor expansion of a function around a point $x_0$ is
\begin{equation}
    f(x) = \sum_{n = 0}^{\infty}\frac{f^{n}(x - x_0)}{n!}(x - x_0)^n.
\end{equation}
Computing the first few derivatives of $g(x)$, it is easy to see that the $n$\textsuperscript{th} derivative of $g(x)$ can be written as $g^{(n)}(x) = x^{1/2 - n}A_n$ where $A_{n+1} = A_{n}(1/2 - n)$ with $A_0 = 1$. This then enables us to reach the closed form for $A_n$ such that for $n\geq 2$, then
\begin{equation}
    A_n = (-1)^{n+1}\frac{(2n-3)!!}{2^n}.
\end{equation}
Thus, for $g(x) = \sqrt{x}$, the closed form equation of the $n^\mathrm{th}$ derivative is written as 
\begin{equation}
    g^{(n)}(x) = x^{1/2 - n}(-1)^{n+1}\frac{(2n-3)!!}{2^n}
\end{equation}
which then can be substituted into the Taylor expansion to compute the any term with $n \geq 2$. The case for $h(x) = \sqrt{\sqrt{x}}$ is done similarly and one arrives at $h^{(n)}(x) = x^{1/4 - n}B_n$ where $B_{n+1} = B_n (1/4 - n)$. It can be seen that for $B_0 = 1$, then $B_1 = 1/4$ and thus for $n\geq 2$, a closed form for the $n^\mathrm{th}$ derivative can be written as
\begin{equation}
    h^{(n)}(x) = x^{1/4 - n}(-1)^{n+1}\frac{(4n-5)!!}{4^n}
\end{equation}
which once again can be substituted into the Taylor expansion of $h(x)$.

\subsection{Random state picking in a Hilbert space}
\label{app:C}

\begin{figure}[ht]
    \centering
    \includegraphics[scale=0.45]{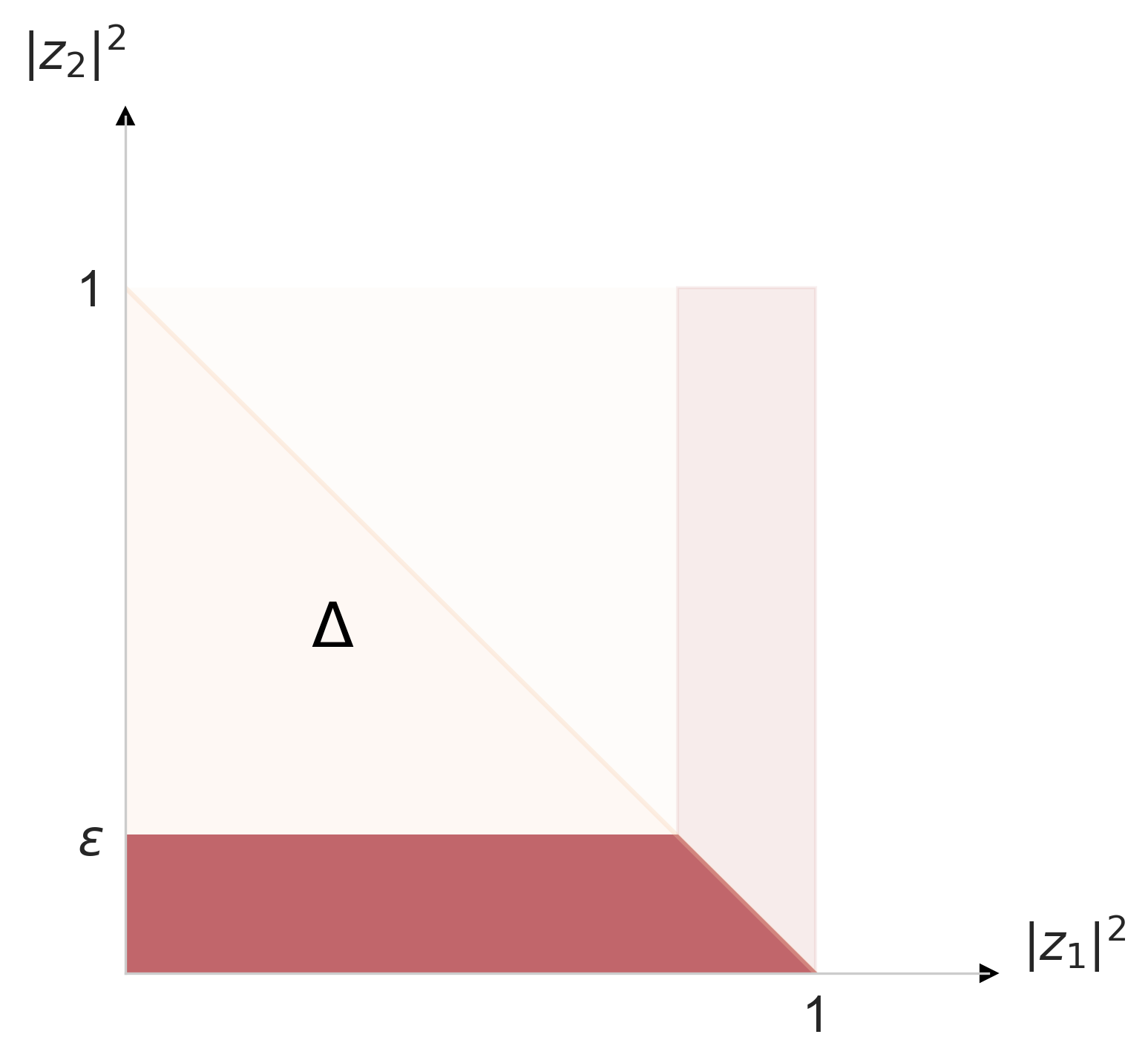}
    \caption{The probability simplex $\Delta$ in a 2-dimensional Hilbert space.}
    \label{fig:simplex}
\end{figure}

To answer the question of what would be observed if the quantum state of a physical system is picked at random from a Hilbert space $\hilb$, we have to specify the probability space. Let $N=\dim \hilb$. The measure is the unique probability measure on the space of unit vectors in $\hilb$ (complex projective space $\mathbb{P}^{N}(\mathbb{C})$) that is invariant under the action of the unitary group U(d). Call $I$ the map 
\begin{equation}
    I:(z_1\ldots z_N) \longmapsto (|z_1|^2, \ldots, |z_N|^2),
\end{equation}
sending a normalized state with coefficients
$z_1\ldots z_N$ to a point in the unit $N$-simplex $\Delta$ (note that $|z_1|^2+\ldots+|z_N|^2=1$).  
Then the pullback of the uniform measure on $\mathbb{P}^{N}$ is the uniform measure on $\Delta$. 
Thus, as long as we are only interested in functions of $|z_1|^2, \ldots, |z_N|^2$, we can work on $\Delta$. The uniform measure on $\Delta$ is given as \cite{Bengtsson_Zyczkowski_2006}
\begin{equation}
\label{eqn:probabilitySimplexMeasure}
    \extd P_\Delta = (N-1)! \, \delta\left(\sum_{i = 1}^N p_i - 1\right) \prod_{i = 1}^{N}\extd p_i.
\end{equation}
The inner product of two quantum states provides insight into the probability amplitude of transitioning from one state to another upon some measurement of the system. To quantify the significance of this transition probability within the Hilbert space, one can consider the process of state selection akin to sampling from the probability simplex formed by the states in the Hilbert space. Moreover, one can quantify the volume of the region within this probability simplex that corresponds to transition probabilities greater than or equal to a certain threshold, which we denote by $\varepsilon$.
\newline
This can be more easily visualised by considering the simple 2-dimensional case in which the probability simplex is as shown in Figure \ref{fig:simplex}. Because of the invariance under unitary transformations of the probability measure, we can assume without loss of generality, that $\Psi$ is not randomly chosen, but is the first basis vector. Then $|\langle \Psi, \Phi\rangle|^2$ corresponds to the random variable $|z_1|^2\equiv p_1$. 
The shaded region correspond to transition probabilities greater than or equal to $\varepsilon$. In this case, then
\begin{align}
    P(|\langle \Psi, \Phi\rangle|^2 \geq \varepsilon) & = 1 - P(|\langle \Psi, \Phi\rangle|^2 < \varepsilon), \\
& = 1 - 2\left(\frac{1 - (1-\varepsilon)^2}{2}\right), \\
& = \varepsilon^2 - 2\varepsilon + 1.
\end{align}
This can be very directly done by computing the area of the square formed by completing the simplex, as shown in the figure, and subtracting from that the area of the square without the red shaded regions and then dividing by two.
\newline
For now arbitrary $N$-dimensions, a general expression can be obtained by first computing the volume of the simplex and the volume of the shaded region. The former is given by dividing the volume of the $N$-dimensional parallelepiped spanned by its edges (which is 1) by $N!$. For the case of the latter, the volume of the region in the simplex bounded by $\varepsilon$ can be found be computing the integral
\begin{align}
    V_\varepsilon(\Delta) & = \int_0^\varepsilon dp_1 \int_0^{1 - p_1}dp_2 \int_0^{1 - p_1 - p_2}dp_3 \dots \int_0^{1 - \sum_{i=1}^{N - 1}p_i}dp_{N} \\
    & = \frac{1 - (1 - \varepsilon)^N}{N!}.
\end{align}
Clearly, $V_\varepsilon(\Delta) \rightarrow 0$ for $\varepsilon \rightarrow 0$ and $V_\varepsilon(\Delta) \rightarrow 1/N!$ for $\varepsilon \rightarrow 1$. Thus, one finds that
\begin{equation}
    P_N(|\langle \Psi, \Phi\rangle|^2 \geq \varepsilon) = 1 - P(|\langle \Psi, \Phi\rangle|^2 < \varepsilon) = 1 - V(\Delta)^{-1} V_\varepsilon(\Delta) = (1 - \varepsilon)^N,
\end{equation}
since $V(\Delta)^{-1} = N!$. This provides a quantitative measure of the likelihood that, given an initial state $\Psi$ within a Hilbert space of $N$-dimensions, a randomly selected state $\Phi$ will exhibit a similarity to $\Psi$ of at least $\varepsilon$. This probability diminishes exponentially with increasing $N$. It can be demonstrated that for $N = 2$, this measure aligns with the previously obtained result above. This is somewhat in analogy to (if one entertains the idea of real valued vectors) the concentration of measure on the sphere in which once a vector $v$ in a unit $N$-sphere is chosen, the probability that second vector chosen at random is orthogonal to it goes to 1 as the dimensions go to infinity \cite{matousek2013lectures}.

\end{document}